\documentclass[aps,prb,twocolumn,superscriptaddress]{revtex4}

\usepackage{graphicx}
\usepackage{bm}

\begin{document}


\title{Modeling the momentum distributions of annihilating
electron-positron pairs in solids}


\author{I. Makkonen}
\email[Electronic address: ]{ima@fyslab.hut.fi}
\affiliation{Laboratory of Physics, Helsinki University of Technology, P.O. Box 1100, FI-02015 HUT, Finland}
\author{M. Hakala}
\affiliation{Division of X-Ray Physics, Department of Physical Sciences, P.O. Box 64, FI-00014 University of Helsinki, Finland}
\author{M. J. Puska}
\affiliation{Laboratory of Physics, Helsinki University of Technology, P.O. Box 1100, FI-02015 HUT, Finland}


\date{\today}

\begin{abstract}
Measuring the Doppler broadening of the positron annihilation radiation 
or the angular correlation between the two annihilation gamma quanta 
reflects the momentum distribution of electrons seen by positrons in the 
material. Vacancy-type defects in solids localize positrons and the 
measured spectra are sensitive to the detailed chemical and geometric
environments of the defects. However, the measured information is 
indirect and when using it in defect identification comparisons with 
theoretically predicted spectra is indispensable.
In this article we present a computational scheme for calculating momentum
distributions of electron-positron pairs annihilating in
solids. Valence electron states and their interaction with ion cores
are
described using the all-electron projector augmented-wave method, and
atomic orbitals are used to describe the core states. We apply our
numerical scheme to selected systems and compare three different
enhancement (electron-positron correlation) schemes previously used
in the calculation of momentum distributions of annihilating
electron-positron pairs within the density-functional theory. We show
that the use of a state-dependent
enhancement scheme leads to better results than a position-dependent
enhancement factor in the case of ratios of Doppler spectra between
different systems. Further, we
demonstrate the applicability of our scheme for studying vacancy-type
defects in metals and semiconductors. Especially we study the
effect of forces due to a positron localized at a vacancy-type
defect on the ionic relaxations.
\end{abstract}


\maketitle


\section{Introduction}

Positron annihilation spectroscopy~\cite{Krause-Rehberg99} is an
experimental method for studying electronic structures of materials.
In comparison with other techniques to measure electron
momentum densities such as $(e,2e)$ spectroscopy~\cite{Coplan94} or
Compton scattering~\cite{Cooper04} (to which we have already
applied the same all-electron method as used in this work~\cite{Makkonen05})
positron annihilation is characterized with the strong sensitivity to the
vacancy-type defects, which makes it a method widely suitable in
materials science and materials technology studies.
In the crystal lattice positrons get trapped at possibly existing
vacancy-type defects. By
measuring positron lifetimes and momentum distributions of annihilating
electron-positron pairs (angular correlation or Doppler broadening
measurements of annihilation radiation) one obtains information about
the open volumes and the chemical environments of the
defects.

A successful use of positron annihilation measurements (in defect
identification) calls for accompanying theoretical and computational
work resulting in simulated annihilation characteristics to be
compared with the measured ones (for a review see Ref.~\onlinecite{Puska94}).
In this paper we present a computational scheme based on the
zero-positron-density ($n_{+}\rightarrow 0$) limit of the
two-component density-functional theory~\cite{Boronski86} (TCDFT). We describe
the valence electron states in materials using the projector
augmented-wave (PAW) method,~\cite{PAW} which we have already used
to simulate the electron momentum distributions measured by Compton
scattering.~\cite{Makkonen05} In the PAW method the core states are
treated in the frozen-core approximation and described using atomic
wave functions.~\cite{Alatalo96} In this work, the positron state is
solved in the real space using a Rayleigh quotient multigrid (RQMG)
solver.~\cite{Heiskanen01} Our scheme gives good results when compared
to experiments for delocalized positron states, for which the
$n_{+}\rightarrow 0$ limit of the TCDFT is exact, as well as for
positrons localized at vacancy-type defects.

The methods previously used in self-consistent
calculations of electronic structures and wave functions for
determination of momentum distributions of annihilating
valence-electron-positron pairs include, for example, the pseudopotential
method,~\cite{Gilgien94,Puska95,Hakala98,Saarinen99,Ishibashi99} the
full-potential linearized-augmented-plane-wave (FLAPW)
method,~\cite{Baruah99,Tang02a,Tang02b} the linear muffin-tin
orbital (LMTO) method~\cite{Barbiellini03} and localized basis
set schemes.~\cite{Ishibashi97} Each of these methods has its own
advantages and flaws. The FLAPW method, although being accurate, is
computationally heavy and has a basis set that makes the momentum
density calculations technically complicated. The pseudopotential
method is efficient and simple to use in the momentum distribution
calculations. The drawback
is that one completely loses the information on the high-momentum
Fourier components of the valence wave functions because soft pseudo
wave functions are used in the calculation. The LMTO method has
presentation problems in the interstitial regions, which renders it
difficult to describe open structures like vacancy-type defects and
systems with low symmetry with it. The PAW
method, in contrast, is as efficient as the ultrasoft pseudopotential
method.~\cite{Vanderbilt90} It can
flexibly be used for the study of defects in solids including
structural relaxation. The plane-wave representation of the pseudo
wave functions makes things simple because plane-waves are eigenfunctions
of the momentum. Moreover, the calculation of the PAW momentum
density~\cite{Makkonen05} is straightforward because the plane-waves
extend also to the augmentation regions as opposed to the (L)APW
method. Ishibashi has already applied the PAW method to the
calculation of coincidence Doppler spectra for bulk
materials,~\cite{Ishibashi04} and Uenodo \textit{et al.}\ have used
the same code to study vacancy-type defects in SiGe.~\cite{Uenodo05}
Previously, we have used the PAW scheme (without taking into
account the effect of the positron-induced forces) to study
vacancy-dopant complexes in highly Sb doped Si,~\cite{Rummukainen05}
monovacancy in Al,~\cite{Calloni05} and to show that one can
experimentally distinguish Ga vacancies (V$_{\text{Ga}}$) from
V$_{\text{Ga}}$--O$_{\text{N}}$ pairs in GaN.~\cite{Hautakangas05}

Many works, in which the interpretation of the results is based on the
comparison between measured and simulated annihilation
characteristics, have been published during the recent
years. Therefore, a systematic study of the performance of different
schemes and approximation is of utmost importance.
Using the PAW method to describe the valence electrons and atomic
orbitals for core electrons we test three different schemes and
approximations to calculate momentum distributions of annihilating
electron-positron pairs.

For the description of the many-body effects
in the calculation of momentum distributions of annihilating
electron-positron pairs we choose finally the so-called state-dependent
scheme~\cite{Alatalo96} and for annihilation rates within the
state-dependent scheme the local-density approximation (LDA)
enhancement factor parametrized by Boro\'nski and
Nieminen.~\cite{Boronski86} We show that use of the commonly used
position-dependent enhancement factor leads to unphysical oscillations
at high momenta when one considers ratios of Doppler
spectra between two different materials. In the same way we compare
the Boro\'nski--Nieminen LDA (BN-LDA) to the generalized-gradient
approximation (GGA) by Barbiellini \textit{et
  al.}~\cite{Barbiellini95,Barbiellini96} Our results show that the
BN-LDA describes the ratios more accurately compared with the
experiment than the GGA\@. We also show that the ratios can be
compared with the experiment reliably although the LDA enhancement
overestimates the high-momentum region of the Doppler spectra arising from
the annihilation with core electrons. There are no test
systems solved theoretically (\textit{e.g.}\ by the Quantum Monte
Carlo method) exactly enough to compare with.

When studying annihilation of positrons trapped at vacancies and
comparing the results with experiments it is important to consider the
effects of forces due to the localized positron on the ionic
relaxation of the vacancy. The effects on calculated positron lifetimes and
Doppler spectra are non-negligible. We study selected monovacancies in
metals and semiconductors by including also the effects of the forces
due to the localized positron.

The structure of the paper is as follows. In section~\ref{theory} we
describe the computational method used. Section~\ref{bulksystems}
presents results for bulk systems and section~\ref{vacancies} selected
results for vacancies in metals and semiconductors. Finally, we
summarize our results and present conclusions in section~\ref{summary}.

\section{Theory and computational methods\label{theory}}

\subsection{Calculation of the positron states}

In our scheme we first calculate the self-consistent electronic
structure of the system without the influence of the positron. Then we
solve the positron state in the potential
\begin{equation}\label{pospot}
V_{+}(\mathbf{r})=\phi(\mathbf{r})+V_{\text{corr}}(n_{-}(\mathbf{r})),
\end{equation}
where $\phi(\mathbf{r})$ is the Coulomb potential due to electrons
and nuclei, $n_{-}(\mathbf{r})$ the electron density, and
$V_{\text{corr}}(n_{-}(\mathbf{r}))$ is the
$n_{+}\rightarrow 0$ limit of the electron-positron correlation
potential. 
Above, the self-interaction
correction is explicitly made, \textit{i.e.}\ since we are interested
in the case of only one positron in the system, its self-direct
Coulomb energy should exactly cancel its exchange-correlation
energy. The calculation of the positron state is non-selfconsistent
because the effective potential for the positron [Eq.~(\ref{pospot})]
does not depend on the positron density.

The scheme described above is for a delocalized positron the exact
$n_{+}\rightarrow 0$ limit of the TCDFT but it has been proven
appropriate in practice also in the case of
localized positron states. In this connection it is often referred
to as the ``conventional scheme''~\cite{Puska94} (CONV). The
approximation can be justified by considering the positron and its
screening cloud as a neutral quasiparticle, which does not affect the
average electron density. One of the difficulties in a full TCDFT
calculation is that the electron-positron correlation functionals are
poorly known at finite positron densities. The $n_{+}\rightarrow 0$
limit used in the CONV scheme is known better.~\cite{Arponen79}

The thermalized positron in the lattice is in the $\mathbf{k}=0$
state. When studying bulk systems we calculate the positron wave
function at the $\Gamma$ point ($\mathbf{k}=0$). In the case of
positrons localized at vacancies the energy eigenvalue corresponding
to an isolated vacancy is broadened due to the supercell approximation
to a narrow band of energies. We integrate over the lowest lying
positron band by calculating the positron wave function both at the
$\Gamma$ point and at the Brillouin zone boundary point $L$ and using
the average of the respective results.~\cite{Korhonen96}

\subsection{Annihilation rate models}

The positron lifetime $\tau$ is the inverse of the annihilation rate $\lambda$
which in a given system is proportional to the overlap of the
electron and positron densities,
\begin{equation}\label{rho}
\lambda=\frac{1}{\tau}=\pi r_{e}^{2}c\int d\mathbf{r}\,n_{-}(\mathbf{r})n_{+}(\mathbf{r})g(n_{-}(\mathbf{r}),n_{+}(\mathbf{r})).
\end{equation}
Above, $r_{e}$ is the classical electron radius and $c$ the speed of
light. The enhancement factor $g(n_{-}(\mathbf{r}),n_{+}(\mathbf{r}))$
(the contact density or the electron-positron pair correlation
function evaluated at the positron) takes into account the increase in the
annihilation due to the screening cloud of electrons around the positron.
[The corresponding result obtained by omitting this factor is called
the independent-particle model (IPM) annihilation rate.] In the TCDFT
and within the LDA $g$ is written as a function of both the local
electron and positron densities. In the CONV
scheme the $n_{+}\rightarrow 0$ limit of the enhancement factor,
denoted by $\gamma(n_{-}(\mathbf{r}))$, is used. Also Gilgien
\textit{et al.}~\cite{Gilgien94} used the $n_{+}\rightarrow 0$ limit
of the enhancement factor in their calculations but they calculated
the positron density self-consistently within the TCDFT\@. This scheme
has been shown to lead to rather localized positron states and too low
core electron annihilation rates in comparison with
experiments.~\cite{Puska95}

The enhancement factor in the Boro\'nski--Nieminen two-component
formalism~\cite{Boronski86} is based on the results of the many-body
calculations by Lantto.~\cite{Lantto87} Gilgien \textit{et
  al.}~\cite{Gilgien94} and Barbiellini \textit{et
al.}~\cite{Barbiellini95,Barbiellini96} have used the
$n_{+}\rightarrow 0$ limit parametrizations consistent with the
correlation energy results of Arponen and Pajanne.~\cite{Arponen79}

The LDA systematically underestimates positron lifetimes in
materials because it overestimates the annihilation with core
electrons for which the correlation effects are less
important.~\cite{Barbiellini95,Barbiellini96} 
Therefore, Barbiellini \textit{et
al.}~\cite{Barbiellini95,Barbiellini96} have presented a
gradient-corrected scheme in which the enhancement factor $\gamma$ is
interpolated between the LDA ($\gamma=\gamma_{\text{LDA}}$, zero
gradient) and the IPM values ($\gamma\equiv 1$, infinite gradient) as a
function of the charge density gradient $\nabla n_{-}$. Effectively,
the interpolation means that the annihilation with valence electrons
in the interstitial region is described using the LDA but when the
density gradient is high (as near nuclei where the rapid oscillations
of core and valence wave functions take place) the enhancement factor
decreases and approaches the IPM limit ($\gamma\equiv 1$). The
interpolation form contains one semi-empirical parameter. The value
$\alpha=0.22$ has been found to give with the Arponen--Pajanne enhancement
lifetimes in good agreement with the
experiment.~\cite{Barbiellini95,Barbiellini96} One must note that also
the correlation potential for the positron is gradient corrected in
the scheme by Barbiellini \textit{et al.} However, the different
enhancement factors cause directly most of the differences compared to the
BN-LDA\@. The LDA parametrization of the correlation potential is the
same in both schemes.

\subsection{Schemes for the calculation of momentum distributions of
  annihilating electron-positron pairs}

The IPM formula for the momentum distribution of annihilating
electron-positron pairs is written as
\begin{equation}
\rho(\mathbf{p})=\pi r_{e}^{2}c\sum_{j}\bigg |\int
d\mathbf{r}\,e^{-i\\\mathbf{p}\cdot\mathbf{r}}\psi_{+}(\mathbf{r})\psi_{j}(\mathbf{r})\bigg
|^{2},
\end{equation}
where $\psi_{+}(\mathbf{r})$ and $\psi_{j}(\mathbf{r})$ are wave
functions of the positron and the electron on orbital $j$,
respectively. The summation goes over the occupied electron
states. The IPM is often used because it gives, in contrast to the
annihilation rate, a rather good qualitative correspondence (shape of
the momentum distribution) with experiments. A common
way to take into account the electron-positron correlation effects is
to introduce in the IPM expression the position-dependent LDA enhancement
factor $\sqrt{\gamma(n_{-}(\mathbf{r}))}$,~\cite{Daniuk87}
\begin{equation}\label{rhoLDA}
\rho(\mathbf{p})=\pi r_{e}^{2}c\sum_{j}\bigg |\int d\mathbf{r}\,e^{-i\mathbf{p}\cdot\mathbf{r}}\psi_{+}(\mathbf{r})\psi_{j}(\mathbf{r})\sqrt{\gamma(n_{-}(\mathbf{r}))}\bigg |^2.
\end{equation}
We call this the state-independent LDA scheme. Eq.~(\ref{rhoLDA}) is,
at least in a homogeneous system,
consistent with the total annihilation rate $\lambda$
of Eq.~(\ref{rho}). Namely, one should obtain $\lambda$ by integrating
over the momentum
\begin{equation}\label{anntot}
\lambda=\int d\mathbf{p}\,\rho(\mathbf{p}).
\end{equation}
The state-independent LDA scheme is motivated by the enhancement
factor of the contact density, but it is not obvious how the screening
really modifies the (electron) wave function. One can consider the
position-dependent enhancement factor
$\sqrt{\gamma(n_{-}(\mathbf{r}))}$ as a factor describing the
distortion of the two-body wave function
$\psi_{+}(\mathbf{r})\psi_{j}(\mathbf{r})$ (where both the electron and
the positron reside at the same point) due to the short-range
electron-positron correlation. What is problematic is that the
two-body wave function is distorted \emph{everywhere} at the instant of
the annihilation although the screening is a local phenomenon. This
causes the correlation effects to be overestimated in the
state-independent LDA scheme.

In the so-called state-dependent scheme~\cite{Alatalo96} a constant
electron-state-dependent enhancement factor $\gamma_{j}$ is used, \textit{i.e.}
\begin{equation}\label{st-dependent}
\rho(\mathbf{p})=\pi r_{e}^{2}c\sum_{j}\gamma_{j}\bigg |\int
d\mathbf{r}\,e^{-i\\\mathbf{p}\cdot\mathbf{r}}\psi_{+}(\mathbf{r})\psi_{j}(\mathbf{r})\bigg
|^{2}. 
\end{equation}
The enhancement factor is written as
$\gamma_{j}=\lambda_{j}/\lambda_{j}^{\text{IPM}}$, where $\lambda_{j}$ is the
annihilation rate of the state $j$ within the LDA or the GGA,
\begin{equation}\label{orbanni}
\lambda_{j}=\pi r_{e}^{2}c\int d\mathbf{r}\, \gamma(n_{-}(\mathbf{r}))n_{+}(\mathbf{r})n_{j}(\mathbf{r}),
\end{equation}
and $\lambda_{j}^{\text{IPM}}$ is the annihilation rate within the
IPM ($\gamma\equiv 1$). Above,
$n_{j}(\mathbf{r})=|\psi_{j}(\mathbf{r})|^{2}$ is the electron density
of the state $j$. In the state-dependent
scheme the momentum
density of a given (electron on a certain orbital) annihilating
electron-positron pair is (apart from the factor $\gamma_{j}$) the
same as in the IPM, \textit{i.e.}\ the enhancement, which in a sense is
averaged over the electron-positron pair, affects only the
annihilation rate $\lambda_{j}$ not the shape of the momentum
distribution of the orbital $j$.
Eq.~(\ref{anntot}) is again satisfied. The problems related to the
state-independent LDA scheme are avoided because the enhancement
factor affects only the probability of annihilation of the positron
with each electron state.

\subsection{Projector augmented-wave method}

\subsubsection{Wave functions in the projector-augmented wave method}

We use the projector augmented-wave (PAW) method~\cite{PAW} to
describe the valence electron wave functions in solids. The PAW method is
a full-potential all-electron method related both to the
pseudopotential method and to the linearized augmented-plane-wave
method (LAPW).
It is based on a linear transformation
between all-electron (AE) valence wave functions $|\Psi\rangle$ and
soft pseudo (PS) valence wave functions $|\tilde{\Psi}\rangle$.
The transformation can be written as (for details see Ref.~\onlinecite{PAW})
\begin{equation}\label{AEwavefunction}
|\Psi\rangle=|\tilde{\Psi}\rangle+\sum_{i}(|\phi_{i}\rangle-|\tilde{\phi}_{i}\rangle)\langle\tilde{p}_{i}|\tilde{\Psi}\rangle,
\end{equation}
where $|\phi_{i}\rangle$ and $|\tilde{\phi}_{i}\rangle$ are AE and PS
partial waves localized around each nucleus, and
$\langle\tilde{p}_{i}|$ are soft, localized projector functions probing
the local character of the PS wave function
$|\tilde{\Psi}\rangle$. Index $i$ stands for the site index $R$,
the angular momentum indices $(l,m)$ and an additional index $k$
referring to the reference energy $\varepsilon_{kl}$.
The solution of the self-consistent electronic structure for a given solid
system means the solution of the PS wave functions. They are represented
by plane-wave expansions in the Vienna \textit{Ab-initio} Simulation
Package~\cite{Kresse96a,Kresse96b,PAWKresse} (\textsc{vasp}) which we
are using. The construction of the AE wave functions in the PAW method is
described in detail in Ref.~\onlinecite{Makkonen05}. The AE valence
wave functions are orthogonal to the core states treated within the
frozen-core approximation (free atom wave functions are used).

When calculating momentum distributions of annihilating
electron-positron pairs, we construct the AE wave functions
$|\Psi\rangle$ according to Eq.~(\ref{AEwavefunction}) in the Fourier
space and then Fourier transform them to the real space. In the case of
positrons localized at defects, the summation over $R$ can be limited
only to the atoms surrounding the defect. The positron state is solved
in the real space. Then the products of the positron and electron
wave functions are calculated and Fourier transformed [see
Eq.~(\ref{st-dependent})]. As a result we have a three-dimensional
momentum distribution on the reciprocal lattice of the
superlattice. Using a dense $\mathbf{k}$-point mesh for electron wave
functions we decrease the lattice constant in order to increase the
momentum resolution and to get a converged result. Then, by integrating
over the planes perpendicular to the chosen momentum distribution the
Doppler spectrum is obtained with a sufficient resolution.

It is sufficient to use a typical value of about 250--400~eV for the
kinetic energy cutoff of the plane-wave expansions when calculating the
PS wave functions for the
determination of the momentum distribution. The momentum components of
the partial waves in Eq.~(\ref{AEwavefunction}) can be taken into
account up to an arbitrary value $p_{\text{max}}$. We have found
that the value $p_{\text{max}}=70\times 10^{-3}\ m_{0}c$ is enough to
guarantee that the Doppler spectrum (projection of $\rho (\mathbf{p})$
on the $p_{z}$ axis) converges up to the momentum of $40\times 10^{-3}\
m_{0}c$, which is usually required when comparing results with coincidence
Doppler broadening experiments.

The PAW method describes also the high-momentum Fourier coefficients of
valence wave functions accurately, which is important when one compares
theoretical results with experimental coincidence Doppler spectra. The
efficiency and the flexibility of the method are also great benefits in
the study of defects in solids. It also enables
one to treat first-row elements, transition metals and rare-earth
elements.

\subsubsection{Constructing the effective potential for the positron}

Although the PAW method is an AE method, we do not in practice
construct the AE valence charge density $n$ in the three-dimensional
real-space grid when constructing the effective potential for the
positron or calculating the total annihilation rate $\lambda$.
A sufficiently good approximation is to approximate $n$ with
$\tilde{n}+\hat{n}$, where
$\tilde{n}$ is the PS valence charge density, calculated from the PS
wave functions $|\tilde{\Psi}\rangle$, and $\hat{n}$ denotes
the compensation charges as defined in
Ref.~\onlinecite{PAWKresse}. (Here we adopt the notation of
Ref.~\onlinecite{PAWKresse}.) The compensation charges $\hat{n}$
guarantee that the approximate Hartree potential due to the valence electrons,
$v_{\text{H}}[\tilde{n}+\hat{n}]$, is equal to the AE Hartree
potential $v_{\text{H}}[n]$ everywhere except near the nuclei,
inside the localized compensation charges $\hat{n}$
($r<r^{l}_{\text{comp}}$, where $r^{l}_{\text{comp}}$'s are the
cutoff radii of the compensation charges). The charge density
$\tilde{n}+\hat{n}$ itself is correct outside the radii
$r_{\text{c}}^{l}$ ($>r^{l}_{\text{comp}}$),
the cutoff radii for the partial
waves $|\phi_{i}\rangle$ and $|\tilde{\phi}_{i}\rangle$, from
nuclei. Typically the radii are, depending on the element, of the order
of $r_{\text{c}}^{l}=1.2\ldots2.3\ a_{0}$ and
$r_{\text{comp}}^{l}=0.8\ldots2.0\ a_{0}$, where $a_{0}$ is the Bohr
radius (see Ref.~\onlinecite{PAWKresse}).

Our approximation is justified by the
fact that near the nuclei the positron density is vanishingly small
because of the Coulomb repulsion of the nuclei. Thereby, the positron
state is not considerably affected and the overlap of the electron and
positron densities (the annihilation rate $\lambda$) does not
appreciably change. Note, however, that we calculate
$n_{j}(\mathbf{r})$, the charge density of the state $j$ in
Eq.~(\ref{orbanni}) represented in the three-dimensional real-space
grid, directly from the AE wave function $\psi_{j}(\mathbf{r})$.

After calculating the Coulomb potential due to both the valence and
core electrons and nuclei, $v_{\text{H}}[\tilde{n}+\hat{n}+n_{Zc}]$,
we calculate the electron-positron correlation potential and solve
the positron state $\psi_{+}(\mathbf{r})$ in the effective potential
$V_{+}(\mathbf{r})$ of Eq.~(\ref{pospot}) using the RQMG
solver.~\cite{Heiskanen01} This is a fast and accurate method for our
purpose where only the positron ground state corresponding to a rather
smooth wave function in the interstitial region has to be solved.

\subsection{Description of positron annihilation with core electrons}

When modeling the positron annihilation and calculating momentum
distributions of annihilating electron-positron pairs we describe the
core electrons and the core electron charge density using atomic
orbitals of isolated atoms calculated within the DFT and the
LDA\@. In the calculation of the electron-positron pair
wave function for the core electron Doppler
spectrum we use an isotropic parametrized positron wave
function~\cite{Alatalo96} of the form
\begin{equation}\label{fit}
\psi_{+}(r)\approx C\{a_{1}+[\text{erf}(r/a_{2})]^{a_{3}}\},
\end{equation}
where $C$ is a normalization factor and $a_{1},a_{2},a_{3}$ are
parameters determined by fitting
Eq.~(\ref{fit}) to a spherically-symmetric positron wave function
calculated with the LMTO method within the atomic-spheres
approximation (ASA).
It is sometimes questionable to assume the positron wave
function to be spherically symmetric around the nuclei when
calculating the core electron Doppler spectrum. 
In the perfect bulk the positron wave function is very isotropic close to
the nuclei. For the positron trapped by a vacancy-type point defect
the decay of the positron wave function in the neighboring ion cores is
similar to that in the bulk. The anisotropy around the nuclei causes
extra localization in the positron-core-electron overlap, which causes
some increase of the positron momentum which is omitted in the
model. However, this is expected to be small in comparison with the
positron momentum due to the decay toward the nuclei.

To test the effects of the frozen-core approximation and the
isotropic positron wavefunction used when modeling the
core-electron Doppler spectrum we have made two calculations
for As vacancy (V$_{\text{As}}$) in GaAs by treating the $3d$ electrons of Ga
first as valence electrons and then as core electrons. In the former
calculation the full three-dimensional positron wave function is used
in constructing the three-dimensional electron-positron pair wave
functions corresponding to the Ga $3d$ electrons whereas in the latter
calculation the positron wave function has the isotropic form of
Eq.~(\ref{fit}). The intensities
of the results differ at high momenta due to the different degree of
self-consistency but when one compares
V$_{\text{As}}$/bulk ratios of Doppler spectra the results coincide. (As long
as the $3d$ electrons of Ga are treated consistently in the bulk and defect
calculations.)

\subsection{Calculation of forces on ions due to a localized positron}

When modeling relaxations of ions around vacancy defects the
effects of the localized positron are included by using the so-called
atomic-superposition (ATSUP) approximation~\cite{Puska94} in which the
charge density and the Coulomb potential are constructed from those of
isolated atoms. In the CONV scheme the total energy is the sum of the
total energy of the electron-ion system and the positron energy
eigenvalue $\varepsilon_{+}$. Then the force due to the positron on ion $j$
is the negative gradient of the positron energy eigenvalue
$\varepsilon_{+}$ with respect to the position of the ion
$\mathbf{R}_{j}$. According to the Hellman--Feynman theorem,
\begin{equation}\label{force}
\mathbf{F}_{j}^{+}=-\nabla_{j}\varepsilon_{+}=-\nabla_{j}\langle\psi^{+}|H|\psi^{+}\rangle=-\langle\psi^{+}|\nabla_{j}H|\psi^{+}\rangle,
\end{equation}
where the positron wave function $|\psi^{+}\rangle$ is assumed to be
properly normalized. Within the ATSUP approximation and the LDA the
effective potential for the positron [Eq.~(\ref{pospot})] is of the form
\begin{equation}\label{pospot2}
V_{+}(\mathbf{r})=\sum_{j}V_{\text{Coul}}^{\text{at},j}(|\mathbf{r}-\mathbf{R}_{j}|)+V_{\text{corr}}\bigg(\sum_{j}n_{-}^{\text{at},j}(|\mathbf{r}-\mathbf{R}_{j}|)\bigg),
\end{equation}
where $V_{\text{Coul}}^{\text{at},j}$ and $n_{-}^{\text{at},j}$
are the Coulomb potential and the charge density of the free atom $j$,
respectively. Inserting this into Eq.~(\ref{force}) gives for the
force
\begin{eqnarray}\label{force2}
\mathbf{F}_{j}^{+} & = & -\int
d\mathbf{r}\,n_{+}(\mathbf{r})\bigg(\frac{\partial
  V^{\text{at},j}_{\text{Coul}}(r)}{\partial r}\bigg|_{|\mathbf{r}-\mathbf{R}_{j}|}\\\nonumber
& + & \frac{\partial V_{\text{corr}}}{\partial  n}\frac{\partial n_{-}^{\text{at},j}(r)}{\partial r}\bigg|_{|\mathbf{r}-\mathbf{R}_{j}|}\bigg)\frac{\mathbf{r}-\mathbf{R}_{j}}{|\mathbf{r}-\mathbf{R}_{j}|}.
\end{eqnarray}

The calculation of the positron-induced forces is fast within the
ATSUP method. The forces are used with calculated electron-ion and
ion-ion forces in order to find the relaxed ionic
configuration of the defect. The ATSUP method itself does not give the
possibility to study the
relaxation of charged defects. However, if the positron state is
solved in the self-consistent Coulomb potential instead of the
superimposed potential of Eq.~(\ref{pospot2}), the effect of the
charge state comes into play, for example, in the case of negatively charged
vacancies in semiconductors, via the stronger localization of the
positron in comparison with the neutral vacancy, which will result in larger
positron-induced forces. The approximations made when using
Eq.~(\ref{force2}) are tested below and compared with TCDFT results.

\section{Perfect bulk systems\label{bulksystems}}

\subsection{Testing the PAW method}

We begin the testing of the PAW method for the calculation of momentum
distributions of annihilating electron-positron pairs by comparing
the results to those of the ATSUP approximation where atomic orbitals
are used also for valence electron states.~\cite{Puska94} Here we
also demonstrate the effect 
of the PAW transformation on the PS wave functions by showing also
results calculated using
only the PS wave functions of the PAW method. In
the calculation of the electronic structures we employ the LDA
exchange-correlation energy. The
positron states and the annihilation characteristics are calculated within the
BN-LDA.~\cite{Boronski86} The Doppler spectra are
calculated using the state-dependent scheme.~\cite{Alatalo96} The
momentum distributions are finally convoluted with a Gaussian function
with a full width at half maximum (FWHM) corresponding to the resolution of
the Doppler experiment.

\begin{figure}[t]
\includegraphics[width=.8\columnwidth]{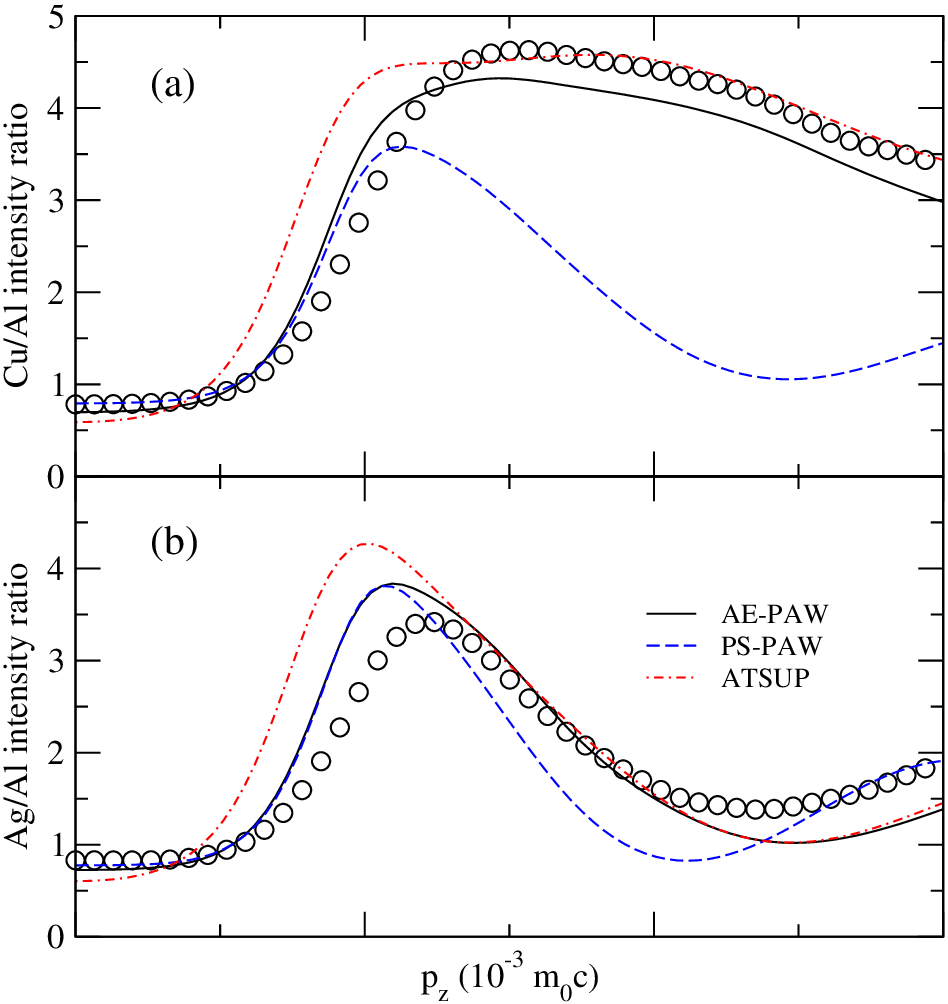}
\caption{(Color online) (a) Bulk Cu / bulk Al and (b) bulk Ag / bulk
  Al ratio curves of
  momentum distributions of annihilating electron-positron
  pairs calculated using the state-dependent scheme. The experimental
  data~\cite{Nagai02} is shown with circles. The theoretical curves
  are convoluted with a Gaussian function with a FWHM of $4.3\times
  10^{-3}\ m_{0}c$.\label{CuAgAl_comparison}}
\end{figure}

We have found out that a good way to compare theoretical and
experimental Doppler spectra is to plot ratios to a reference
spectrum. This way most of the systematic error is canceled. For the
theoretical spectra this means especially that the overestimation of
the core annihilation in the LDA is not a problem. In
Fig.~\ref{CuAgAl_comparison} we show our
results for the ratios Cu/Al and Ag/Al calculated with the AE-PAW method,
the PS wave functions of the PAW method and the ATSUP method compared
to the experimental data from Ref.~\onlinecite{Nagai02}. In these
calculations the Cu $3d$ and Ag $4d$ electrons are treated as valence
electrons in the PAW calculations.

Due to the non-selfconsistent construction of the valence electron
wave functions the ATSUP method gives good results only at high momenta
where the annihilation with the core electrons dominates. The results
obtained with the AE-PAW method are clearly better in the low-momentum
region of the spectra. In the high-momentum region these two results
are equally good. The better compatibility of the crude ATSUP
approximation with the experiment at high momenta in
Fig.~\ref{CuAgAl_comparison}(a) may be just a coincidence. The PS wave
functions of the PAW method fail to
represent the high-frequency oscillations of the valence wave
functions, especially those of the Cu $3d$ and Ag $4d$ electrons, in
the core region although they predict the ratios well up to
the momentum of about $10\times 10^{-3}\ m_{0}c$.
The AE-PAW results in Fig.~\ref{CuAgAl_comparison}
are practically combinations of the ATSUP (at high
momenta) and the PS-PAW (at low momenta) results. Because the quality
of the PS-PAW results is comparable to calculations made using
norm-conserving pseudowave functions~\cite{Hakala98,Saarinen99} the
tests made in this section clearly show the benefits of the use of the
PAW method in the accurate calculation of momentum distributions of
annihilating electron-positron pairs. We also note that positron lifetimes
obtained for different bulk systems agree perfectly with previous
all-electron results calculated with the same enhancement factor.


\subsection{State-dependent scheme vs.\ state-independent LDA scheme}

In this section we compare the two above-mentioned ways to take into
account the electron-positron correlation in the calculation of the
momentum distribution of annihilating electron-positron pairs: the
state-dependent scheme of Eq.~(\ref{st-dependent}) and the state-independent
LDA scheme of Eq.~(\ref{rhoLDA}). For simplicity, we use only bulk
materials (Cu, Ag, Al, Si, Mo, and Fe) as examples. We use the
experimental lattice constants and for Fe the experimental bcc
structure. From now on we use the AE-PAW method within the frozen-core
approximation.

\begin{figure}[t]
\includegraphics[width=.8\columnwidth]{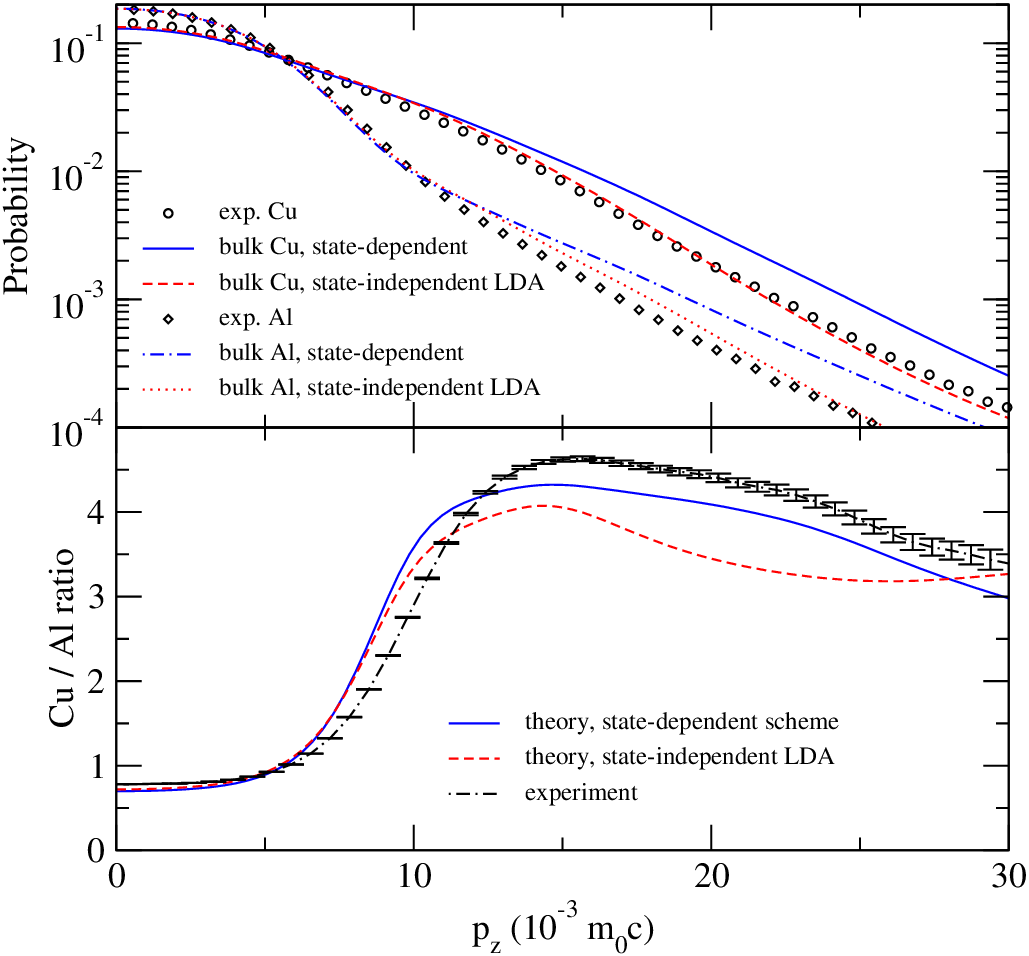}
\caption{(Color online) Bulk Cu / bulk Al ratio curve of momentum
  distributions of annihilating electron-positron pairs. The comparison
  between the state-dependent and the state-independent LDA schemes is
  shown. The
  experimental data is from Ref.~\onlinecite{Nagai02}. The
  theoretical curves are convoluted with a Gaussian function with a
  FWHM of $4.3\times 10^{-3}\ m_{0}c$.\label{CuAl}}
\end{figure}

\begin{figure}[t]
\includegraphics[width=.8\columnwidth]{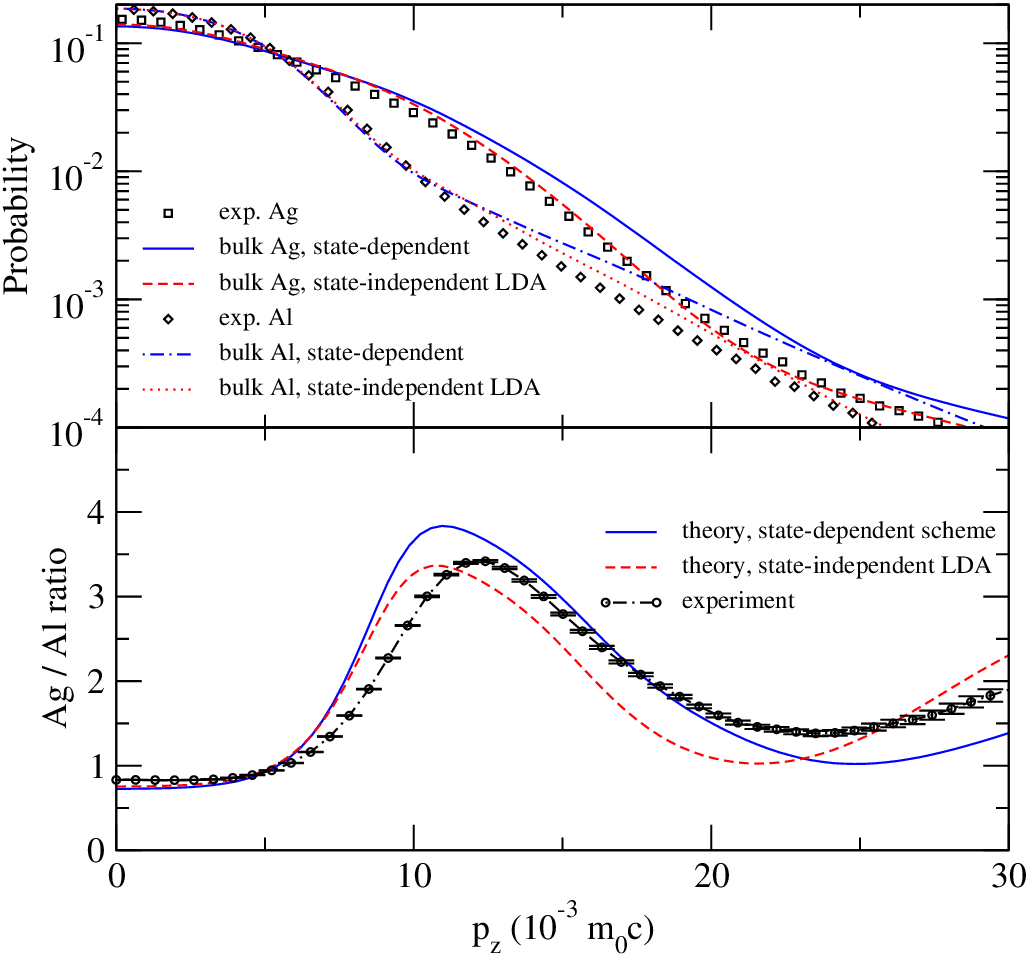}
\caption{(Color online) Bulk Ag / bulk Al ratio curve of momentum
  distributions 
  of annihilating electron-positron pairs. The comparison between the
  state-dependent and the state-independent LDA schemes is shown. The
  experimental data is from Ref.~\onlinecite{Nagai02}. The
  theoretical curves are convoluted with a Gaussian function with a
  FWHM of $4.3\times 10^{-3}\ m_{0}c$.\label{AgAl}}
\end{figure}

We show in Figs.~\ref{CuAl} and~\ref{AgAl} the
Doppler spectra of Cu, Ag and Al calculated within the both
above-mentioned schemes in the logarithmic
scale and also normalize the Doppler spectra to the one
of Al. The theoretical spectra and ratios are compared with the
experimental ones by Nagai \textit{et al}.~\cite{Nagai02} Both schemes
describe the low-momentum region due to the annihilation with valence
electrons well but the ratios tell, as seen
before,~\cite{Alatalo96} that the state-independent LDA scheme fails to
describe the high-momentum region of the Doppler spectra which arises
from the annihilation with core electrons. The ratio Cu/Al (see
Fig.~\ref{CuAl}) calculated using the
state-independent LDA scheme is not even in a qualitative agreement
with the experiment at high momenta. On the contrary, the
state-dependent scheme describes the ratios quite accurately in both
Figs.~\ref{CuAl} and~\ref{AgAl}.
However, when one looks at the absolute values
of the spectra, the intensities at high momenta are in better agreement
with the experiment in the results calculated using the
state-independent LDA scheme but there is some unphysical oscillation
in the spectra that does not exist in the state-dependent LDA results.

\begin{figure}[t]
\includegraphics[width=.8\columnwidth]{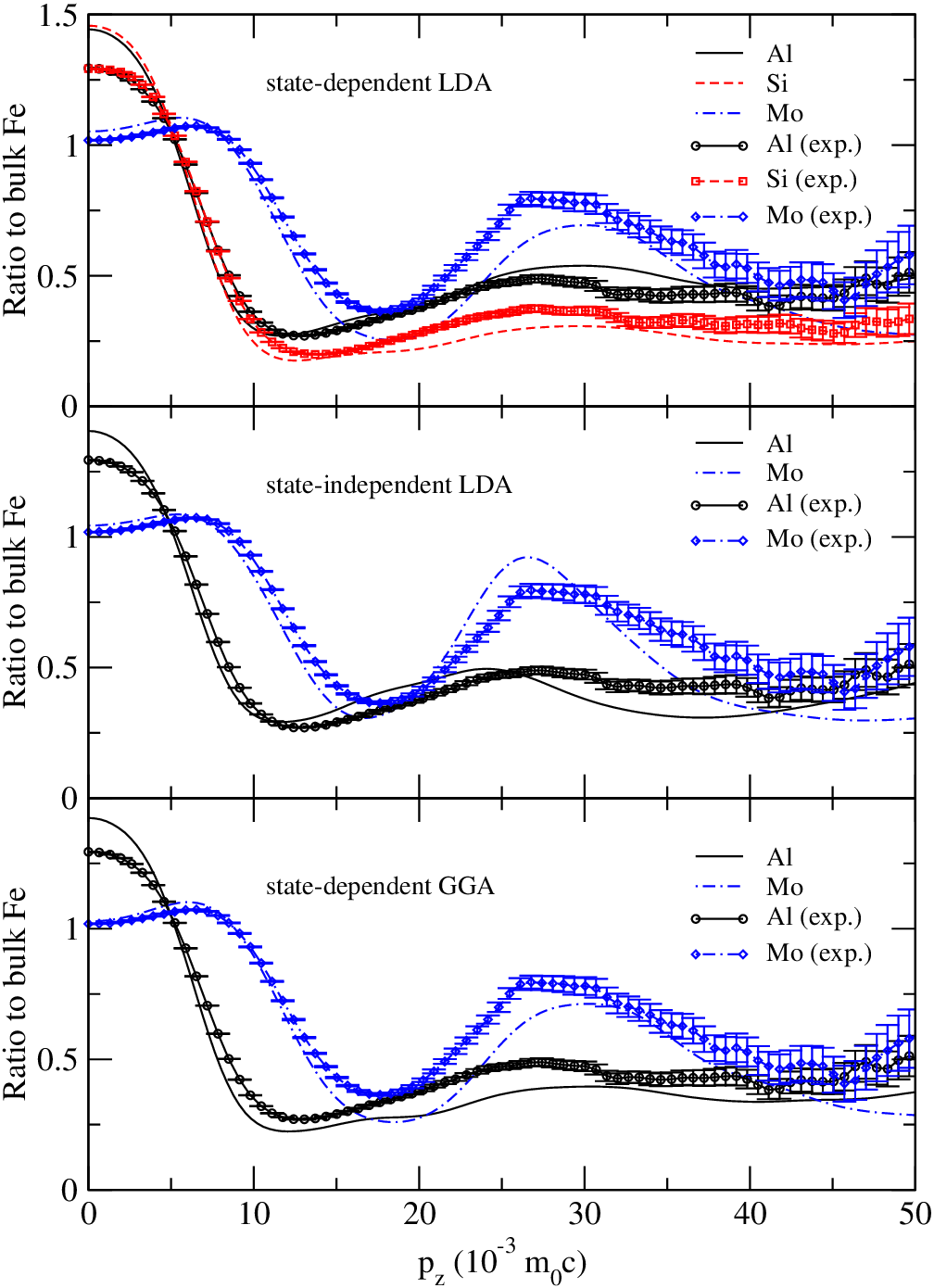}
\caption{(Color online) Ratio curves of momentum distributions of annihilating
  electron-positron pairs calculated using different
  approximations. The experimental data is from
  Ref.~\onlinecite{Nagai00}. The theoretical curves are convoluted
  with a Gaussian function with a FWHM of $4.7\times 10^{-3}\
  m_{0}c$.\label{AlSiMoFe}}
\end{figure}

\begin{figure}[t]
\includegraphics[width=.8\columnwidth]{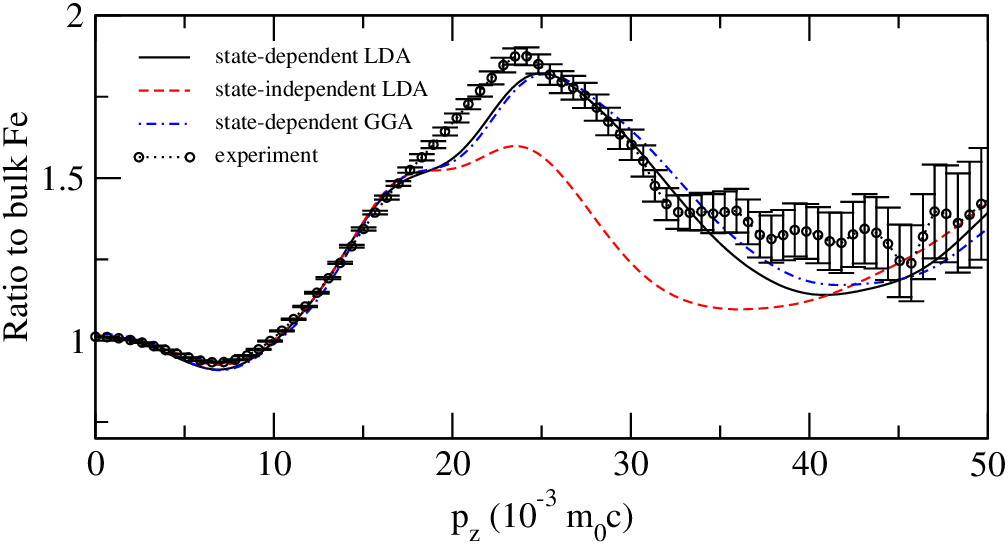}
\caption{(Color online) Bulk Cu / bulk Fe ratio curve of momentum
  distributions of annihilating
  electron-positron pairs calculated using different
  approximations. The experimental data is from
  Ref.~\onlinecite{Nagai00}. The theoretical curves are convoluted
  with a Gaussian function with a FWHM of $4.7\times 10^{-3}\
  m_{0}c$.\label{CuFe}}
\end{figure}

Further examples are shown in Figs.~\ref{AlSiMoFe} and~\ref{CuFe} where
the Doppler spectra of Al, Si, Mo and Cu are normalized to the one for Fe. For
Fe we consider its magnetic ground state. The state-independent LDA
fails again at high momenta, the result for Cu being again in the
worst agreement with the experiment.

\subsection{LDA vs.\ GGA within the state-dependent scheme}

\begin{figure}[t]
\includegraphics[width=.8\columnwidth]{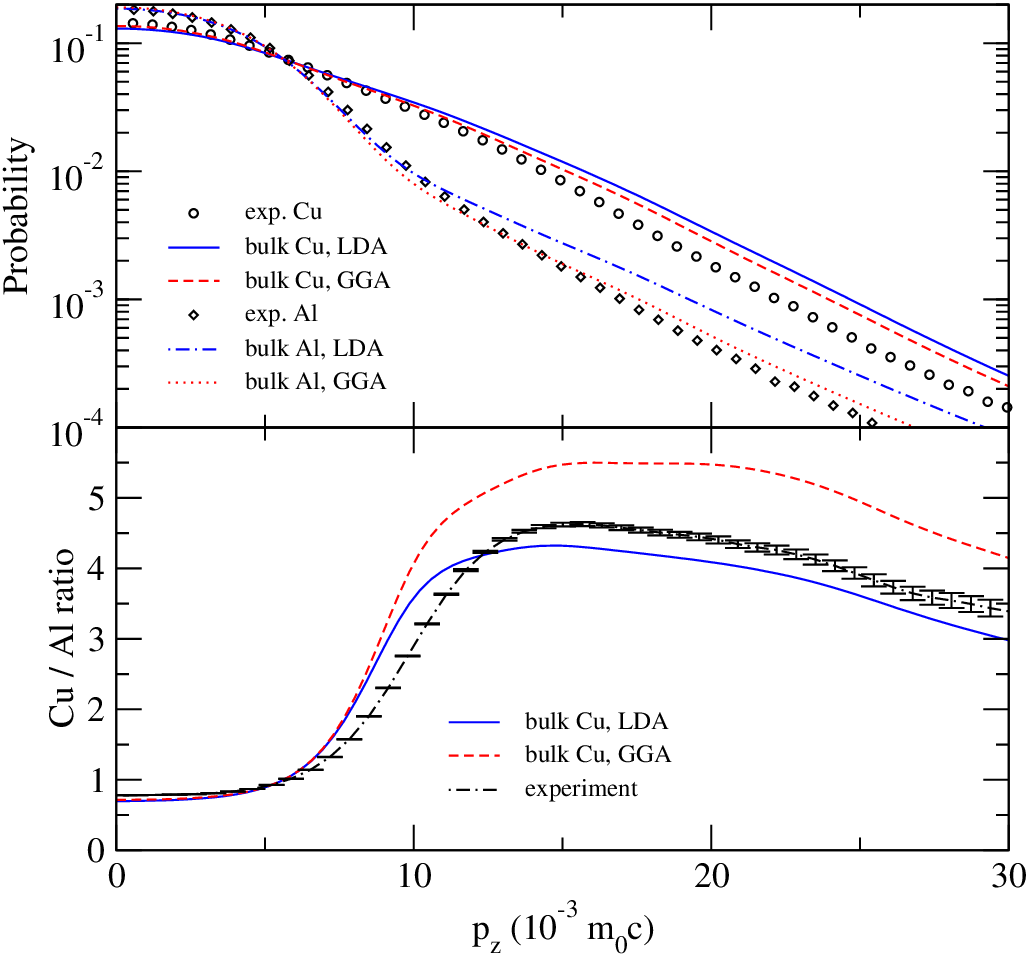}
\caption{(Color online) Bulk Cu / bulk Al ratio curve of momentum
  distributions
  of annihilating electron-positron pairs. The comparison between the
  BN-LDA and the GGA for annihilation rates within the state-dependent
  scheme is shown. The
  experimental data is from Ref.~\onlinecite{Nagai02}. The
  theoretical curves are convoluted with a Gaussian function with a
  FWHM of $4.3\times 10^{-3}\ m_{0}c$.\label{CuAl_LDA_GGA}}
\end{figure}

\begin{figure}[t]
\includegraphics[width=.8\columnwidth]{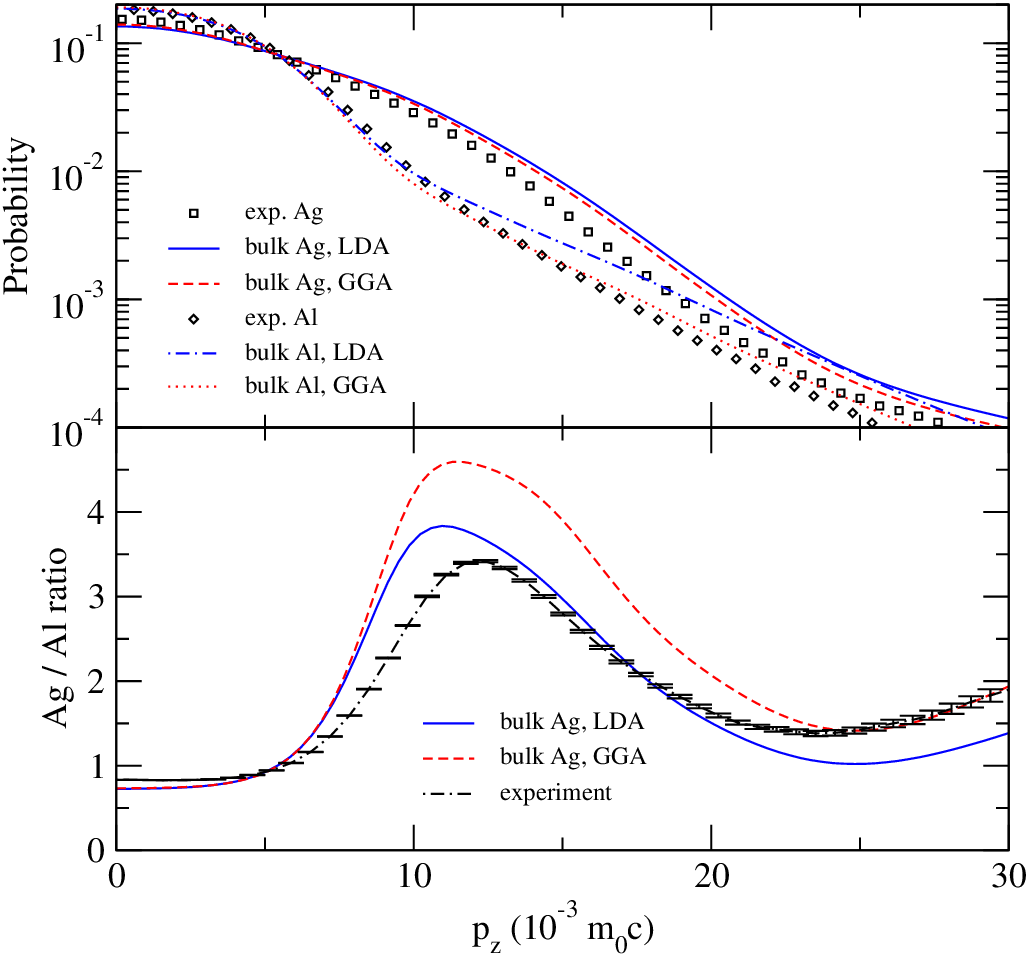}
\caption{(Color online) Bulk Ag / bulk Al ratio curve of momentum
  distributions
  of annihilating electron-positron pairs. The comparison between the
  BN-LDA and the GGA for annihilation rates within the
  state-dependent scheme is shown. The
  experimental data is from Ref.~\onlinecite{Nagai02}. The
  theoretical curves are convoluted with a Gaussian function with a
  FWHM of $4.3\times 10^{-3}\ m_{0}c$.\label{AgAl_LDA_GGA}}
\end{figure}

In Figs.~\ref{AlSiMoFe}, \ref{CuFe}, \ref{CuAl_LDA_GGA},
and~\ref{AgAl_LDA_GGA} we show similar comparisons between the BN-LDA
and the GGA by Barbiellini \textit{et
  al.}~\cite{Barbiellini95,Barbiellini96} The state-dependent scheme
is used. The results calculated using the BN-LDA for the annihilation
rates are in a better agreement
with the experiment. In Figs.~\ref{CuAl_LDA_GGA}
and~\ref{AgAl_LDA_GGA} the GGA tends to give too high values for the
ratio at high momenta. In these particular examples the failure is
mainly due to the large decrease of the relative core annihilation
rate of Al calculated with the GGA compared to the one according to
the BN-LDA\@. Because the state-dependent scheme is used the shapes of
the contributions due to individual orbitals are the same and also
the shapes of the ratios are very similar.
The ratios Al/Fe, Mo/Fe and Cu/Fe calculated with the GGA are shown in
Figs.~\ref{AlSiMoFe} and~\ref{CuFe}. The results are in a rather good
agreement with the experiment and only slightly worse than the ones
calculated with the BN-LDA state-dependent scheme. The
relative underestimation of the core annihilation in Al can be seen
also in the ratio Al/Fe, which is lower at high momenta than the
BN-LDA result. In contrast, as seen in Figs.~\ref{CuAl_LDA_GGA}
and~\ref{AgAl_LDA_GGA} the GGA describes better the \textit{absolute}
intensities of the Doppler spectra because the annihilation rates of
core orbitals are decreased.

The failure of the GGA in the ratios of Doppler spectra can be traced
back to the semiempirical interpolation form of the GGA enhancement
factor. Although its zero- and high-gradient limits are well defined
the interpolation form is only an approximation. Moreover, the free
parameter $\alpha$ is just fixed to give \emph{lifetimes} that are in good
agreement with the experiment. Clearly, the BN-LDA succeeds to
describe better the relative magnitudes of the annihilation rates
$\lambda_{j}$ between different electronic states and different
elements.

We conclude that our choice for the approximation to be used with the
state-dependent scheme is the BN-LDA because it gives better results
than the GGA when comparing intensity ratios with the experiment. It
is also simpler and more justifiable.

\section{Vacancy defects in solids\label{vacancies}}

The following step is to demonstrate that our scheme works also for
positrons localized at vacancy-type defects. The CONV scheme
has been shown to yield for a given ionic structure
lifetimes~\cite{Korhonen96} and also other
annihilation characteristics~\cite{Puska95} in good agreement with
two-component calculations based on the Boro\'nski--Nieminen
formalism.

In this section we
compare our results for vacancies in metals and semiconductors to experimental
Doppler broadening results. We also study the effect of the
positron-induced forces on ions neighboring
vacancies. Further, we compare the ionic relaxations to previous
two-component results and compare Doppler spectra calculated with the
relaxed structures with experimental ones.

\subsection{Relaxation tests}

We study the effect of the localized positron on the relaxation of
ions surrounding a vacancy by calculating the ionic structures of
monovacancies in bulk Si, Al and Cu with and without the localized
positron. 
We consider first only
isotropic relaxations. For Si we use a cubic 64-atom supercell and for
Al and Cu cubic 108-atom supercells. In the electronic structure
calculations we sample the Brillouin zone
using $4^{3}$,  $8^{3}$ and $6^{3}$ Monkhorst--Pack~\cite{Monkhorst76}
$\mathbf{k}$-point meshes, respectively. Self-consistent LDA lattice
constants are used in all calculations. We use the value 0.01~eV/\AA\ as
a stopping criterion for the forces on ions when finding the
self-consistent ionic configurations.

We consider two different approximations for the force calculation. We
solve the positron state either in the PAW potential or in the ATSUP
potential. The first approximation is better in the sense that the
positron density is able to follow the changes in the electronic
structure but our force expression, Eq.~(\ref{force2}), is based on the
ATSUP approximation and therefore the PAW potential is not consistent
with the potential of the force calculation and thus the total energy
minimum does not correspond to vanishing forces. In contrast, the
latter more crude approximation is consistent with the force
expression used and gives thus a well-defined and consistent total
energy minimum.

\begin{figure}
\includegraphics[width=.75\columnwidth]{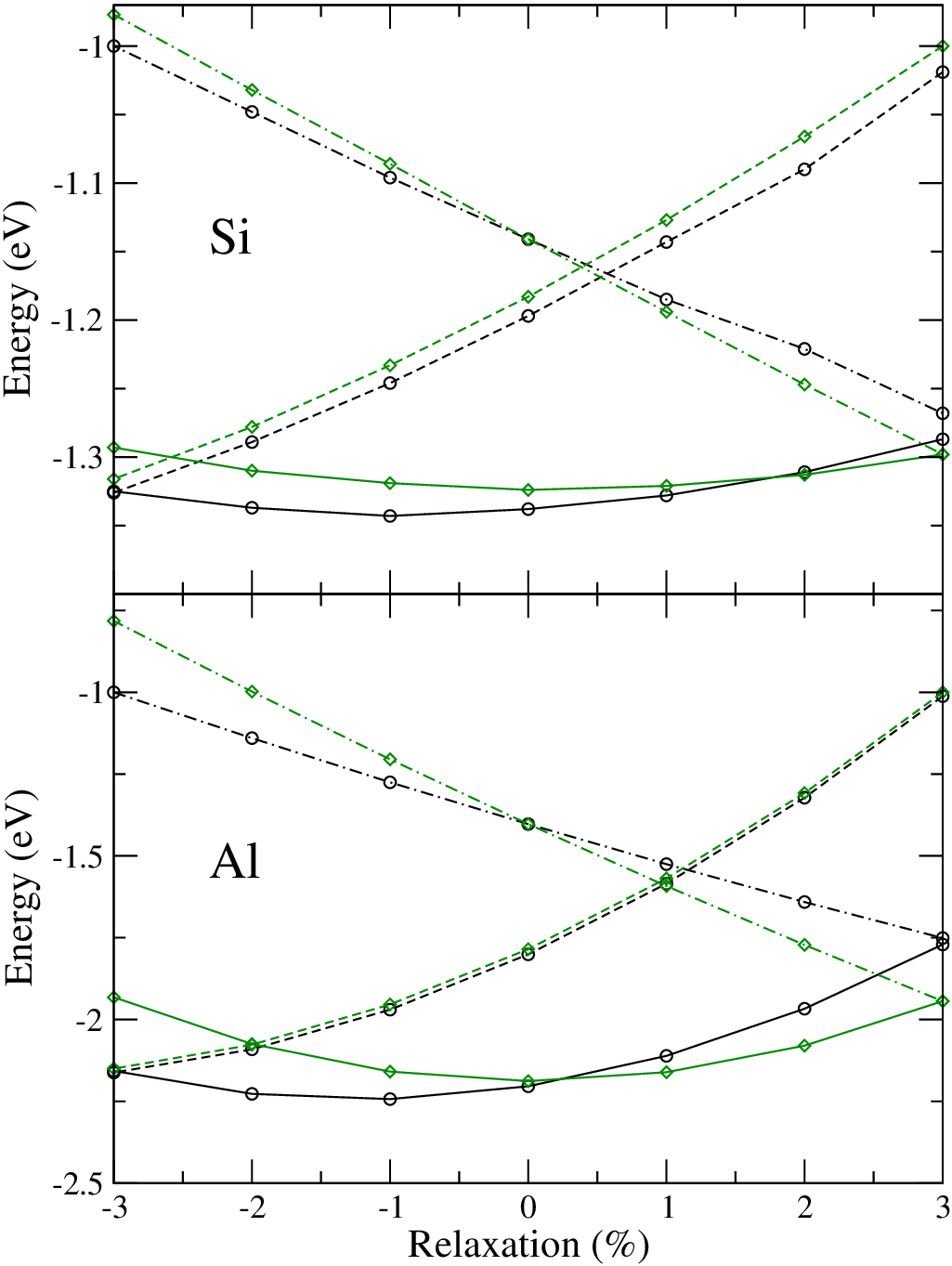}
\caption{(Color online) Different energy components as functions of vacancy
  relaxation in \% of the nearest-neighbor distance (relative to the
  relaxed geometry, only nearest neighbor ions are moved) for V$_{\text{Si}}$
  in Si and V$_{\text{Al}}$ in Al. The solid
  lines denote the total energy, the dashed lines the energy of the
  electron-ion system, and the dash-dotted line the positron energy
  eigenvalue. The figures show results in which the forces are
  calculated using a positron state solved in the PAW
  potential~($\circ$) and in the ATSUP
  potential~($\diamond$).\label{rel}}
\end{figure}

The relaxations obtained with and without the positron are listed in
Table~\ref{relaxations}. The effect of the positron on the relaxations
is clear; in all of these examples the inward relaxation is
transformed to an outward relaxation due to the positron. The
relaxations obtained with the PAW and the ATSUP potentials are very
similar. We have studied the Si vacancy with the localized positron
using also a larger cubic 216-atom supercell and only the $\Gamma$
point. The results obtained are shown
in Table~\ref{relaxations} in parenthesis. The larger outward
relaxation is explained by the fact that in the larger supercell the
ions can relax more freely; interactions between periodic
images of the vacancy are not as dominant as in the 64-atom
supercell. In Fig.~\ref{rel} we
have plotted different energy components
of the V$_{\text{Si}}$ and V$_{\text{Al}}$ systems as
functions of vacancy relaxation relative to the relaxed
geometry (ionic structure obtained including the effect of the forces
due to the positron). Note, that one can not compare absolute energies
between the results calculated with different potentials for the
positron. The energy minimum for the PAW positron potential plus
potential due to the electron-ion system is in both
systems at about 1~\% smaller relaxation than the structure given by
the relaxation (zero forces).

\begin{table}
\caption{Relaxations and corresponding positron lifetimes $\tau$ for
  monovacancies in different bulk
  materials. A positive (negative) number denotes the isotropic
  outward (inward) relaxation in percentage of the nearest neighbor
  distance with
  respect to the unrelaxed (ideal) vacancy. The table
  includes results calculated without the effect of the positron and
  with the positron state solved in the PAW/ATSUP potential. The
  relaxation is restricted to the symmetric breathing-mode
  relaxation. The numbers
  in parenthesis are calculated using a larger 216-atom
  supercell. Computed bulk lifetimes are 208~ps, 159~ps and 95~ps for
  Si, Al and Cu, respectively.\label{relaxations}}
\begin{ruledtabular}
\begin{tabular}{ccccccc}
& \multicolumn{2}{c}{no $e^{+}$} & \multicolumn{2}{c}{PAW} &
\multicolumn{2}{c}{ATSUP}\\
& rel.\ (\%) & $\tau$ (ps) & rel.\ (\%) & $\tau$ (ps) & rel.\ (\%) &
  $\tau$ (ps)\\
\hline
Si & --10.4 & 215 & +5.6 (+11.3) & 256 (272) & +5.9 & 257\\
Al & --1.7 & 219 & +2.8 & 242 & +2.9 & 251\\
Cu &  --1.3 & 146 & +2.4 & 163\\
\end{tabular}
\end{ruledtabular}
\end{table}

Our computational positron lifetimes for the monovacancies in Al and
Cu are in good agreement with the experiment. The experimental
lifetimes for V$_{\text{Al}}$ and V$_{\text{Cu}}$ are
251~ps (Ref.~\onlinecite{Schaefer84}) and 179~ps
(Ref.~\onlinecite{Schaefer86}), respectively. They are high in
comparison with the computed lifetimes since we use the LDA enhancement
factor but we find a
good agreement when we compare the vacancy-bulk lifetime differences
with the experiment. (The experimental lifetimes for bulk Al and bulk
Cu are 170~ps and 120~ps, respectively.~\cite{Barbiellini96})
The calculated and experimental results for the monovacancy in Si are
compared in section~\ref{VSi}.

\subsection{Ga vacancy in GaAs}

The triply negative Ga vacancy in GaAs has been extensively studied by
Puska \textit{et al.}~\cite{Puska95} using the TCDFT and different schemes
(including the CONV) for the electron-positron
correlation. Furthermore, also experimental coincidence Doppler
broadening data exists.~\cite{Laine96} Thus, using the Ga vacancy as a
benchmark system, it is possible to compare simultaneously the
relaxations and lifetimes obtained to two-component results and
lifetimes and Doppler
spectra to the experiment. We model the GaAs lattice using a cubic
64-atom supercell and sample the Brillouin zone in the
electronic-structure calculations with a $4^{3}$
Monkhorst--Pack $\mathbf{k}$-point mesh.~\cite{Monkhorst76}
In the case of the triply
negative charge state of V$_{\text{Ga}}$ all the localized states in the band
gap are occupied and there is no symmetry lowering Jahn--Teller
relaxation. By including the effect of the positron-induced forces on the
relaxation we get an inward relaxation of 5.9~\% in very good agreement
with the inward relaxation of 6.6~\% previously obtained using the
two-component BN scheme.~\cite{Puska95} In the work by Puska
\textit{et al.}~\cite{Puska95}\ the authors did not calculate
relaxations using the CONV scheme but showed that the total energy
curves calculated as a function of the breathing-mode relaxations of
the ions neighboring the vacancy using the the two-component BN and
the CONV schemes nearly coincide. When this and the good agreement in
relaxations are taken into account, one can conclude that our scheme
for the calculation of forces gives very similar results to ones
calculated using the two-component BN formalism.

\begin{figure}[t]
\includegraphics[width=.8\columnwidth]{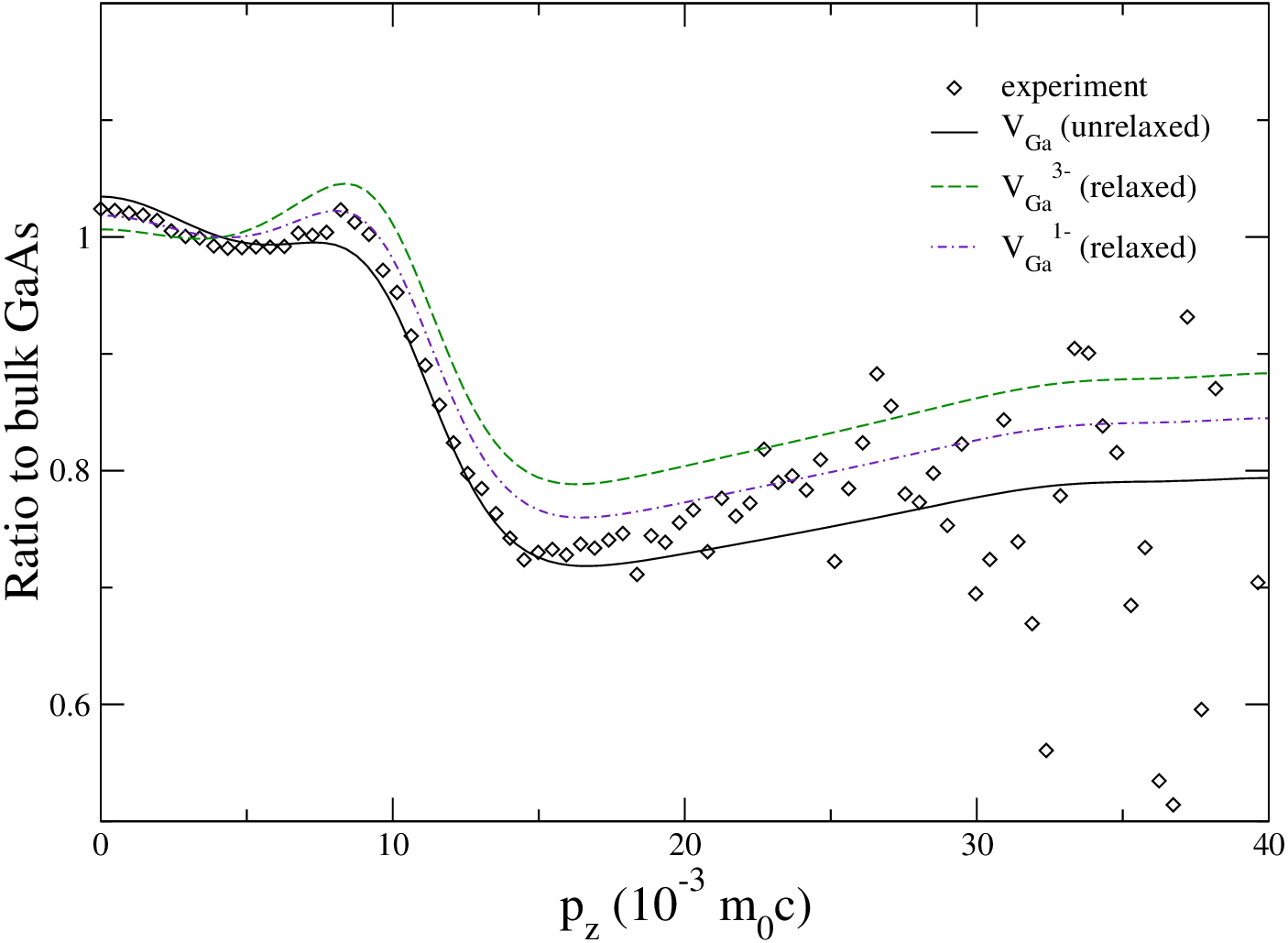}
\caption{(Color online) Theoretical ratio curves for the Ga vacancy in
  GaAs. The
  experimental data, measured from electron-irradiated GaAs, is from
  Ref.~\onlinecite{Laine96}. The theoretical curves are convoluted with a Gaussian function with a FWHM of $5.5\times 10^{-3}\ m_{0}c$.\label{VGa}}
\end{figure}

In Fig.~\ref{VGa} we show the computed Doppler spectra
(normalized to that of bulk GaAs) obtained for the relaxed
structures of V$_{\text{Ga}}$ in the charge states $3-$ and $1-$ compared
with the experiment and the computed one for the neutral, ideal
(unrelaxed) V$_{\text{Ga}}$. The corresponding relaxations and lifetimes are
tabulated in Table~\ref{VGa_lifetimes}. Only isotropic relaxations
have been considered using the small 64-atom supercell although a
symmetry-breaking relaxation is expected for the $1-$ charge
state. The agreement with the experiment both in the Doppler spectrum
and in the lifetime (relative to the bulk one) is best for the $1-$
charge state. The inward relaxation for the $3-$ is too strong
compared with the experiment because the lifetime and the ratio curve
in Fig.~\ref{VGa} are too close to the bulk values.

\begin{table}
\caption{Relaxations and lifetimes for different charge states of
  V$_{\text{Ga}}$ in GaAs. The numbers in parenthesis are calculated using a
  larger 216-atom supercell. The relaxation is restricted to the
  symmetric breathing-mode relaxation. A positive (negative) number
  denotes isotropic outward (inward) relaxation. The computed bulk
  GaAs lifetime is 208~ps.\label{VGa_lifetimes}}
\begin{ruledtabular}
\begin{tabular}{cccc}
defect & rel.\ (\%) & $\tau$ (ps) & $\tau-\tau_{\text{bulk}}$ (ps)\\
\hline
unrel.\ V$_{\text{Ga}}^{0}$ & 0.0 & 249 & 38\\
V$_{\text{Ga}}^{1-}$ & --2.8 (--2.3) & 237 (244) & 26 (33)\\
V$_{\text{Ga}}^{3-}$ & --5.9 & 229 & 18\\
V$_{\text{Ga}}$ (exp.~\cite{Corbel92}) & & 260 & 30\\
\end{tabular}
\end{ruledtabular}
\end{table}

We have also calculated the relaxation of V$_{\text{Ga}}^{1-}$ using
a larger cubic 216-atom supercell (with this supercell we use only the
$\Gamma$ point and do not treat the Ga $3d$ electrons as
valence electrons). The calculation gives slightly smaller inward relaxation
than that with the 64-atom cell. The results are given in
Table~\ref{VGa_lifetimes} in parenthesis. We also break the $T_{d}$
symmetry by displacing the nearest-neighbor atoms of the vacancy in
order to create a symmetry-breaking relaxation with the expected
$C_{2v}$ symmetry. The corresponding lifetime
is 243~ps, which is almost the same as when assuming the T$_{d}$
symmetry. (Note that Fig.~\ref{VGa} does not include any results calculated
with the 216-atom supercell.)

The relative core annihilation rate (the experimentally measured relative $W$
parameter, relative wrt bulk value) is sensitive to the treatment of the
electron-positron correlation effects. The correlation potential used
affects the degree of the localization of the positron and thereby the
positron-core electron overlap and the core annihilation
rate.~\cite{Puska95} 
Puska \textit{et al.}~\cite{Puska95} obtained
for the triply negative Ga vacancy in GaAs the values of 0.88 and 0.34
using the two-component BN formalism and the scheme
by Gilgien \textit{et al}.,~\cite{Gilgien94} respectively. The
computational values in
the present work (estimated from Fig.~\ref{VGa}) are of the order of
0.8 in good agreement with the experiment and the previous result
obtained with the BN formalism. Our scheme has already previously been
successful in describing the relative core annihilation rates in the
case of Ga vacancy in GaN~\cite{Hautakangas05} and monovacancy in
Al.~\cite{Calloni05}


\subsection{Neutral monovacancy in Si\label{VSi}}

The neutral monovacancy in Si is a very complex system with a flat
potential-energy surface and several competing local
minima as a function of ion
positions.~\cite{Puska98,Probert03} One can get even qualitatively
different results with two different approximations \textit{e.g.}\ for
the exchange-correlation potential.~\cite{Probert03} We study now the
monovacancy in Si including the forces caused by the localized
positron. We use the 216-atom supercell and the $\Gamma$ point, begin from
the structure obtained as a result from the isotropic relaxation (see
Table~\ref{VGa_lifetimes}) and break the $T_{d}$ symmetry by
displacing the nearest-neighbor atoms of the vacancy in order to
create a symmetry-breaking relaxation with the expected $D_{2d}$
symmetry. We confirm the fact pointed out by several earlier
studies~\cite{Puska98,Probert03,Makhov05,Latham05} that a supercell with at
least 216 atoms is needed in order to obtain a convergence with
respect to the supercell size.

\begin{figure}[t]
\includegraphics[width=.7\columnwidth]{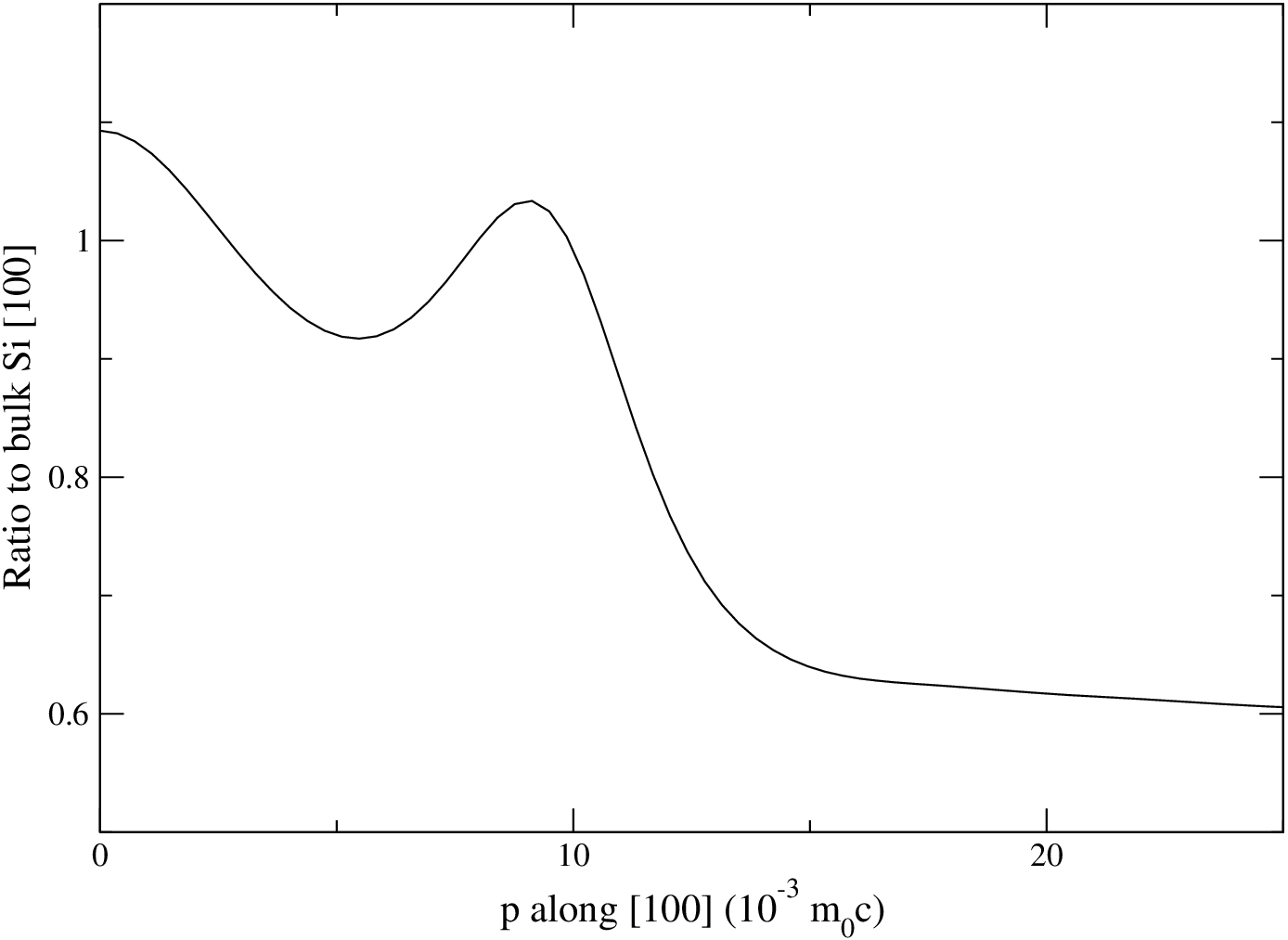}
\caption{Theoretical ratio curve for the neutral Si vacancy in Si. The
  the data is convoluted with a Gaussian function with a FWHM of
  $3.7\times 10^{-3}\ m_{0}c$\label{ratio_VSi}}
\end{figure}

Our calculation gives an outward-relaxed structure with a slight
$D_{2d}$ symmetry. The corresponding lifetime is 270~ps; only 2~ps
less than that for the vacancy constrained to the $T_{d}$ symmetry. In
Fig.~\ref{ratio_VSi} we show the obtained vacancy/bulk ratio of
Doppler spectra. No experimental Doppler broadening data exists for
the monovacancy in Si. M\"akinen \textit{et al.}~\cite{Makinen89} and
Polity \textit{et al.}~\cite{Polity98} obtained the positron lifetimes
of 273 and 282~ps, respectively, for the monovacancy in Si created by electron
irradiation. Taking into account that the calculated bulk lifetime with
the LDA lattice constant used (208~ps) is lower than the experimental
ones by M\"akinen \textit{et al.}\ and Polity \textit{et al.}\ (221
and 218~ps, respectively) we conclude that our result for the lifetime
is in a good agreement with experiment.

Previously, Saito and Oshiyama~\cite{Saito96} and Makhov and
Lewis~\cite{Makhov05} have studied vacancies
in Si within the two-component scheme by Gilgien \textit{et
  al}.~\cite{Gilgien94} including the effects of forces due to the
positron. Their lifetimes for the monovacancy are reasonable but the
relative $W$ parameter of 0.28 estimated from Saito and Oshiyama's data is
very low in comparison with our result, which is 0.72 
(both evaluated using calculated annihilation rates as in
  Ref.~\onlinecite{Puska95}).

\section{Summary and conclusions\label{summary}}

In conclusion, we have presented an accurate scheme for the
calculation of momentum distributions of annihilating
electron-positron pairs in solids based on the projector
augmented-wave method.

We have compared three commonly used approaches for the momentum
distributions within the DFT framework.
We have shown that the most appropriate way to take into account screening
effects in the calculation of momentum distributions is to use a
constant, state-dependent enhancement factor. Further, we have demonstrated
that a position-dependent enhancement factor gives unphysical results
when ratios of Doppler spectra are considered. The differences in
results of the BN-LDA and the GGA by
Barbiellini \textit{et al.}\ is not large. Except for one of the
studied elements (Al) the GGA gives comparable results. We choose
to rely on the BN-LDA because of its simplicity and
lack of semi-empiric parameters. We underline that our choices are
based on the theory-experiment comparison of ratios of momentum
distributions for different materials rather than their absolute
values. The latter may often be better described in the GGA and in the
position-dependent enhancement schemes.

In addition to bulk solids our scheme is also reliable in the case of
defects in materials. The comparison of our results for the Ga vacancy
in GaAs to experiment suggests that the Ga vacancy seen in the experiment
is negative but less than triply negative, which is
the charge state suggested by a recent \textit{ab initio}
study.~\cite{El-Mellouhi05} For the neutral monovacancy in Si we have
presented a prediction to be compared with future lifetime and
coincidence Doppler broadening experiments.

\begin{acknowledgments}
We thank Prof.\ M.\ Hasegawa and Prof.\ Y.\ Nagai for providing the
experimental coincidence Doppler broadening data from the bulk
materials and Prof.\ K.\ Saarinen for discussions and for the GaAs
data. We thank also Prof.\ H.\ Yukawa for performing preliminary
positron calculations for bulk metals. We also acknowledge the
generous computer resources from the Center of Scientific Computing,
Espoo, Finland. This work has been
supported by the Academy of Finland through its Centers of Excellence program
(2000-2005) and contract No.\ 201291/205967 (MH), by Research Funds of
the University of Helsinki (MH), and by the Finnish Academy of Science and
Letters, Vilho, Yrj\"o and Kalle V\"ais\"al\"a Foundation (IM).
\end{acknowledgments}

\bibliography{positronpaper}

\begin{thebibliography}{49}
\expandafter\ifx\csname natexlab\endcsname\relax\def\natexlab#1{#1}\fi
\expandafter\ifx\csname bibnamefont\endcsname\relax
  \def\bibnamefont#1{#1}\fi
\expandafter\ifx\csname bibfnamefont\endcsname\relax
  \def\bibfnamefont#1{#1}\fi
\expandafter\ifx\csname citenamefont\endcsname\relax
  \def\citenamefont#1{#1}\fi
\expandafter\ifx\csname url\endcsname\relax
  \def\url#1{\texttt{#1}}\fi
\expandafter\ifx\csname urlprefix\endcsname\relax\def\urlprefix{URL }\fi
\providecommand{\bibinfo}[2]{#2}
\providecommand{\eprint}[2][]{\url{#2}}

\bibitem[{\citenamefont{Krause-Rehberg and Leipner}(1999)}]{Krause-Rehberg99}
\bibinfo{author}{\bibfnamefont{R.}~\bibnamefont{Krause-Rehberg}}
  \bibnamefont{and} \bibinfo{author}{\bibfnamefont{H.}~\bibnamefont{Leipner}},
  \emph{\bibinfo{title}{Positron Annihilation in Semiconductors}}
  (\bibinfo{publisher}{Springer-Verlag}, \bibinfo{address}{Berlin},
  \bibinfo{year}{1999}).

\bibitem[{\citenamefont{Coplan et~al.}(1994)\citenamefont{Coplan, Moore, and
  Doering}}]{Coplan94}
\bibinfo{author}{\bibfnamefont{M.~A.} \bibnamefont{Coplan}},
  \bibinfo{author}{\bibfnamefont{J.~H.} \bibnamefont{Moore}}, \bibnamefont{and}
  \bibinfo{author}{\bibfnamefont{J.~P.} \bibnamefont{Doering}},
  \bibinfo{journal}{Rev.\ Mod.\ Phys.} \textbf{\bibinfo{volume}{66}},
  \bibinfo{pages}{985} (\bibinfo{year}{1994}).

\bibitem[{\citenamefont{Cooper et~al.}(2004)\citenamefont{Cooper, Mijnarends,
  Shiotani, Sakai, and Bansil}}]{Cooper04}
\bibinfo{editor}{\bibfnamefont{M.~J.} \bibnamefont{Cooper}},
  \bibinfo{editor}{\bibfnamefont{P.~E.} \bibnamefont{Mijnarends}},
  \bibinfo{editor}{\bibfnamefont{N.}~\bibnamefont{Shiotani}},
  \bibinfo{editor}{\bibfnamefont{N.}~\bibnamefont{Sakai}}, \bibnamefont{and}
  \bibinfo{editor}{\bibfnamefont{A.}~\bibnamefont{Bansil}}, eds.,
  \emph{\bibinfo{title}{X-Ray Compton Scattering}} (\bibinfo{publisher}{Oxford
  University Press}, \bibinfo{address}{Oxford}, \bibinfo{year}{2004}).

\bibitem[{\citenamefont{Makkonen et~al.}(2005)\citenamefont{Makkonen, Hakala,
  and Puska}}]{Makkonen05}
\bibinfo{author}{\bibfnamefont{I.}~\bibnamefont{Makkonen}},
  \bibinfo{author}{\bibfnamefont{M.}~\bibnamefont{Hakala}}, \bibnamefont{and}
  \bibinfo{author}{\bibfnamefont{M.~J.} \bibnamefont{Puska}},
  \bibinfo{journal}{J.\ Phys.\ Chem.\ Solids} \textbf{\bibinfo{volume}{66}},
  \bibinfo{pages}{1128} (\bibinfo{year}{2005}).

\bibitem[{\citenamefont{Puska and Nieminen}(1994)}]{Puska94}
\bibinfo{author}{\bibfnamefont{M.~J.} \bibnamefont{Puska}} \bibnamefont{and}
  \bibinfo{author}{\bibfnamefont{R.~M.} \bibnamefont{Nieminen}},
  \bibinfo{journal}{Rev. Mod. Phys.} \textbf{\bibinfo{volume}{66}},
  \bibinfo{pages}{841} (\bibinfo{year}{1994}).

\bibitem[{\citenamefont{Boro\'nski and Nieminen}(1986)}]{Boronski86}
\bibinfo{author}{\bibfnamefont{E.}~\bibnamefont{Boro\'nski}} \bibnamefont{and}
  \bibinfo{author}{\bibfnamefont{R.~M.} \bibnamefont{Nieminen}},
  \bibinfo{journal}{Phys. Rev. B} \textbf{\bibinfo{volume}{34}},
  \bibinfo{pages}{3820} (\bibinfo{year}{1986}).

\bibitem[{\citenamefont{Bl\"ochl}(1994)}]{PAW}
\bibinfo{author}{\bibfnamefont{P.~E.} \bibnamefont{Bl\"ochl}},
  \bibinfo{journal}{Phys. Rev. B} \textbf{\bibinfo{volume}{50}},
  \bibinfo{pages}{17953} (\bibinfo{year}{1994}).

\bibitem[{\citenamefont{Alatalo et~al.}(1996)\citenamefont{Alatalo,
  Barbiellini, Hakala, Kauppinen, Korhonen, Puska, Saarinen, Hautoj\"arvi, and
  Nieminen}}]{Alatalo96}
\bibinfo{author}{\bibfnamefont{M.}~\bibnamefont{Alatalo}},
  \bibinfo{author}{\bibfnamefont{B.}~\bibnamefont{Barbiellini}},
  \bibinfo{author}{\bibfnamefont{M.}~\bibnamefont{Hakala}},
  \bibinfo{author}{\bibfnamefont{H.}~\bibnamefont{Kauppinen}},
  \bibinfo{author}{\bibfnamefont{T.}~\bibnamefont{Korhonen}},
  \bibinfo{author}{\bibfnamefont{M.~J.} \bibnamefont{Puska}},
  \bibinfo{author}{\bibfnamefont{K.}~\bibnamefont{Saarinen}},
  \bibinfo{author}{\bibfnamefont{P.}~\bibnamefont{Hautoj\"arvi}},
  \bibnamefont{and} \bibinfo{author}{\bibfnamefont{R.~M.}
  \bibnamefont{Nieminen}}, \bibinfo{journal}{Phys. Rev. B}
  \textbf{\bibinfo{volume}{54}}, \bibinfo{pages}{2397} (\bibinfo{year}{1996}).

\bibitem[{\citenamefont{Heiskanen et~al.}(2001)\citenamefont{Heiskanen, Torsti,
  Puska, and Nieminen}}]{Heiskanen01}
\bibinfo{author}{\bibfnamefont{M.}~\bibnamefont{Heiskanen}},
  \bibinfo{author}{\bibfnamefont{T.}~\bibnamefont{Torsti}},
  \bibinfo{author}{\bibfnamefont{M.~J.} \bibnamefont{Puska}}, \bibnamefont{and}
  \bibinfo{author}{\bibfnamefont{R.~M.} \bibnamefont{Nieminen}},
  \bibinfo{journal}{Phys. Rev. B} \textbf{\bibinfo{volume}{63}},
  \bibinfo{pages}{245106} (\bibinfo{year}{2001}).

\bibitem[{\citenamefont{Saarinen et~al.}(1999)\citenamefont{Saarinen,
  Nissil\"a, Kauppinen, Hakala, Puska, Hautoj\"arvi, and Corbel}}]{Saarinen99}
\bibinfo{author}{\bibfnamefont{K.}~\bibnamefont{Saarinen}},
  \bibinfo{author}{\bibfnamefont{J.}~\bibnamefont{Nissil\"a}},
  \bibinfo{author}{\bibfnamefont{H.}~\bibnamefont{Kauppinen}},
  \bibinfo{author}{\bibfnamefont{M.}~\bibnamefont{Hakala}},
  \bibinfo{author}{\bibfnamefont{M.~J.} \bibnamefont{Puska}},
  \bibinfo{author}{\bibfnamefont{P.}~\bibnamefont{Hautoj\"arvi}},
  \bibnamefont{and} \bibinfo{author}{\bibfnamefont{C.}~\bibnamefont{Corbel}},
  \bibinfo{journal}{Phys. Rev. Lett.} \textbf{\bibinfo{volume}{82}},
  \bibinfo{pages}{1883} (\bibinfo{year}{1999}).

\bibitem[{\citenamefont{Gilgien et~al.}(1994)\citenamefont{Gilgien, Galli,
  Gygi, and Car}}]{Gilgien94}
\bibinfo{author}{\bibfnamefont{L.}~\bibnamefont{Gilgien}},
  \bibinfo{author}{\bibfnamefont{G.}~\bibnamefont{Galli}},
  \bibinfo{author}{\bibfnamefont{F.}~\bibnamefont{Gygi}}, \bibnamefont{and}
  \bibinfo{author}{\bibfnamefont{R.}~\bibnamefont{Car}},
  \bibinfo{journal}{Phys.\ Rev.\ Lett.} \textbf{\bibinfo{volume}{72}},
  \bibinfo{pages}{3214} (\bibinfo{year}{1994}).

\bibitem[{\citenamefont{Puska et~al.}(1995)\citenamefont{Puska, Seitsonen, and
  Nieminen}}]{Puska95}
\bibinfo{author}{\bibfnamefont{M.~J.} \bibnamefont{Puska}},
  \bibinfo{author}{\bibfnamefont{A.~P.} \bibnamefont{Seitsonen}},
  \bibnamefont{and} \bibinfo{author}{\bibfnamefont{R.~M.}
  \bibnamefont{Nieminen}}, \bibinfo{journal}{Phys.\ Rev.\ B}
  \textbf{\bibinfo{volume}{52}}, \bibinfo{pages}{10947} (\bibinfo{year}{1995}).

\bibitem[{\citenamefont{Hakala et~al.}(1998)\citenamefont{Hakala, Puska, and
  Nieminen}}]{Hakala98}
\bibinfo{author}{\bibfnamefont{M.}~\bibnamefont{Hakala}},
  \bibinfo{author}{\bibfnamefont{M.~J.} \bibnamefont{Puska}}, \bibnamefont{and}
  \bibinfo{author}{\bibfnamefont{R.~M.} \bibnamefont{Nieminen}},
  \bibinfo{journal}{Phys. Rev. B} \textbf{\bibinfo{volume}{57}},
  \bibinfo{pages}{7621} (\bibinfo{year}{1998}).

\bibitem[{\citenamefont{Ishibashi et~al.}(1999)\citenamefont{Ishibashi, Manuel,
  Kohyama, Tokumoto, and Anzai}}]{Ishibashi99}
\bibinfo{author}{\bibfnamefont{S.}~\bibnamefont{Ishibashi}},
  \bibinfo{author}{\bibfnamefont{A.~A.} \bibnamefont{Manuel}},
  \bibinfo{author}{\bibfnamefont{M.}~\bibnamefont{Kohyama}},
  \bibinfo{author}{\bibfnamefont{M.}~\bibnamefont{Tokumoto}}, \bibnamefont{and}
  \bibinfo{author}{\bibfnamefont{H.}~\bibnamefont{Anzai}},
  \bibinfo{journal}{Phys.\ Rev.\ B} \textbf{\bibinfo{volume}{60}},
  \bibinfo{pages}{R3747} (\bibinfo{year}{1999}).

\bibitem[{\citenamefont{Baruah et~al.}(1999)\citenamefont{Baruah, Zope, and
  Kshirsagar}}]{Baruah99}
\bibinfo{author}{\bibfnamefont{T.}~\bibnamefont{Baruah}},
  \bibinfo{author}{\bibfnamefont{R.~R.} \bibnamefont{Zope}}, \bibnamefont{and}
  \bibinfo{author}{\bibfnamefont{A.}~\bibnamefont{Kshirsagar}},
  \bibinfo{journal}{Phys.\ Rev. B} \textbf{\bibinfo{volume}{60}},
  \bibinfo{pages}{10770} (\bibinfo{year}{1999}).

\bibitem[{\citenamefont{Tang et~al.}(2002{\natexlab{a}})\citenamefont{Tang,
  Hasegawa, Nagai, Saito, and Kawazoe}}]{Tang02a}
\bibinfo{author}{\bibfnamefont{Z.}~\bibnamefont{Tang}},
  \bibinfo{author}{\bibfnamefont{M.}~\bibnamefont{Hasegawa}},
  \bibinfo{author}{\bibfnamefont{Y.}~\bibnamefont{Nagai}},
  \bibinfo{author}{\bibfnamefont{M.}~\bibnamefont{Saito}}, \bibnamefont{and}
  \bibinfo{author}{\bibfnamefont{Y.}~\bibnamefont{Kawazoe}},
  \bibinfo{journal}{Phys.\ Rev. B} \textbf{\bibinfo{volume}{65}},
  \bibinfo{pages}{045108} (\bibinfo{year}{2002}{\natexlab{a}}).

\bibitem[{\citenamefont{Tang et~al.}(2002{\natexlab{b}})\citenamefont{Tang,
  Hasegawa, Nagai, and Saito}}]{Tang02b}
\bibinfo{author}{\bibfnamefont{Z.}~\bibnamefont{Tang}},
  \bibinfo{author}{\bibfnamefont{M.}~\bibnamefont{Hasegawa}},
  \bibinfo{author}{\bibfnamefont{Y.}~\bibnamefont{Nagai}}, \bibnamefont{and}
  \bibinfo{author}{\bibfnamefont{M.}~\bibnamefont{Saito}},
  \bibinfo{journal}{Phys. Rev. B} \textbf{\bibinfo{volume}{65}},
  \bibinfo{pages}{195108} (\bibinfo{year}{2002}{\natexlab{b}}).

\bibitem[{\citenamefont{Barbiellini et~al.}(2003)\citenamefont{Barbiellini,
  Dugdale, and Jarlborg}}]{Barbiellini03}
\bibinfo{author}{\bibfnamefont{B.}~\bibnamefont{Barbiellini}},
  \bibinfo{author}{\bibfnamefont{S.~B.} \bibnamefont{Dugdale}},
  \bibnamefont{and} \bibinfo{author}{\bibfnamefont{T.}~\bibnamefont{Jarlborg}},
  \bibinfo{journal}{Comput.\ Mater.\ Sci.} \textbf{\bibinfo{volume}{28}},
  \bibinfo{pages}{287} (\bibinfo{year}{2003}).

\bibitem[{\citenamefont{Ishibashi et~al.}(1997)\citenamefont{Ishibashi, Manuel,
  Hoffmann, and Bechgaard}}]{Ishibashi97}
\bibinfo{author}{\bibfnamefont{S.}~\bibnamefont{Ishibashi}},
  \bibinfo{author}{\bibfnamefont{A.~A.} \bibnamefont{Manuel}},
  \bibinfo{author}{\bibfnamefont{L.}~\bibnamefont{Hoffmann}}, \bibnamefont{and}
  \bibinfo{author}{\bibfnamefont{K.}~\bibnamefont{Bechgaard}},
  \bibinfo{journal}{Phys.\ Rev.\ B} \textbf{\bibinfo{volume}{55}},
  \bibinfo{pages}{2048} (\bibinfo{year}{1997}).

\bibitem[{\citenamefont{Vanderbilt}(1990)}]{Vanderbilt90}
\bibinfo{author}{\bibfnamefont{D.}~\bibnamefont{Vanderbilt}},
  \bibinfo{journal}{Phys. Rev. B} \textbf{\bibinfo{volume}{41}},
  \bibinfo{pages}{R7892} (\bibinfo{year}{1990}).

\bibitem[{\citenamefont{Ishibashi}(2004)}]{Ishibashi04}
\bibinfo{author}{\bibfnamefont{S.}~\bibnamefont{Ishibashi}},
  \bibinfo{journal}{Mater.\ Sci.\ Forum} \textbf{\bibinfo{volume}{445-446}},
  \bibinfo{pages}{401} (\bibinfo{year}{2004}).

\bibitem[{\citenamefont{Uenodo et~al.}(2005)\citenamefont{Uenodo, Hattori,
  Naruoka, Ishibashi, Suzuki, and Ohdaira}}]{Uenodo05}
\bibinfo{author}{\bibfnamefont{A.}~\bibnamefont{Uenodo}},
  \bibinfo{author}{\bibfnamefont{N.}~\bibnamefont{Hattori}},
  \bibinfo{author}{\bibfnamefont{H.}~\bibnamefont{Naruoka}},
  \bibinfo{author}{\bibfnamefont{S.}~\bibnamefont{Ishibashi}},
  \bibinfo{author}{\bibfnamefont{R.}~\bibnamefont{Suzuki}}, \bibnamefont{and}
  \bibinfo{author}{\bibfnamefont{T.}~\bibnamefont{Ohdaira}},
  \bibinfo{journal}{J.\ Appl.\ Phys.} \textbf{\bibinfo{volume}{97}},
  \bibinfo{pages}{023532} (\bibinfo{year}{2005}).

\bibitem[{\citenamefont{Rummukainen et~al.}(2005)\citenamefont{Rummukainen,
  Makkonen, Ranki, Puska, Saarinen, and Gossmann}}]{Rummukainen05}
\bibinfo{author}{\bibfnamefont{M.}~\bibnamefont{Rummukainen}},
  \bibinfo{author}{\bibfnamefont{I.}~\bibnamefont{Makkonen}},
  \bibinfo{author}{\bibfnamefont{V.}~\bibnamefont{Ranki}},
  \bibinfo{author}{\bibfnamefont{M.~J.} \bibnamefont{Puska}},
  \bibinfo{author}{\bibfnamefont{K.}~\bibnamefont{Saarinen}}, \bibnamefont{and}
  \bibinfo{author}{\bibfnamefont{H.-J.~L.} \bibnamefont{Gossmann}},
  \bibinfo{journal}{Phys.\ Rev.\ Lett.} \textbf{\bibinfo{volume}{94}},
  \bibinfo{pages}{165501} (\bibinfo{year}{2005}).

\bibitem[{\citenamefont{Calloni et~al.}(2005)\citenamefont{Calloni, Dupasquier,
  Ferragut, Folegati, Iglesias, Makkonen, and Puska}}]{Calloni05}
\bibinfo{author}{\bibfnamefont{A.}~\bibnamefont{Calloni}},
  \bibinfo{author}{\bibfnamefont{A.}~\bibnamefont{Dupasquier}},
  \bibinfo{author}{\bibfnamefont{R.}~\bibnamefont{Ferragut}},
  \bibinfo{author}{\bibfnamefont{P.}~\bibnamefont{Folegati}},
  \bibinfo{author}{\bibfnamefont{M.~M.} \bibnamefont{Iglesias}},
  \bibinfo{author}{\bibfnamefont{I.}~\bibnamefont{Makkonen}}, \bibnamefont{and}
  \bibinfo{author}{\bibfnamefont{M.~J.} \bibnamefont{Puska}},
  \bibinfo{journal}{Phys.\ Rev.\ B} \textbf{\bibinfo{volume}{72}},
  \bibinfo{pages}{054112} (\bibinfo{year}{2005}).

\bibitem[{\citenamefont{Hautakangas et~al.}(2005)\citenamefont{Hautakangas,
  Makkonen, Ranki, Puska, Saarinen, Xu, and Look}}]{Hautakangas05}
\bibinfo{author}{\bibfnamefont{S.}~\bibnamefont{Hautakangas}},
  \bibinfo{author}{\bibfnamefont{I.}~\bibnamefont{Makkonen}},
  \bibinfo{author}{\bibfnamefont{V.}~\bibnamefont{Ranki}},
  \bibinfo{author}{\bibfnamefont{M.~J.} \bibnamefont{Puska}},
  \bibinfo{author}{\bibfnamefont{K.}~\bibnamefont{Saarinen}},
  \bibinfo{author}{\bibfnamefont{X.}~\bibnamefont{Xu}}, \bibnamefont{and}
  \bibinfo{author}{\bibfnamefont{D.~C.} \bibnamefont{Look}}
  (\bibinfo{year}{2005}), \bibinfo{note}{to be published}.

\bibitem[{\citenamefont{Barbiellini et~al.}(1995)\citenamefont{Barbiellini,
  Puska, Torsti, and Nieminen}}]{Barbiellini95}
\bibinfo{author}{\bibfnamefont{B.}~\bibnamefont{Barbiellini}},
  \bibinfo{author}{\bibfnamefont{M.~J.} \bibnamefont{Puska}},
  \bibinfo{author}{\bibfnamefont{T.}~\bibnamefont{Torsti}}, \bibnamefont{and}
  \bibinfo{author}{\bibfnamefont{R.~M.} \bibnamefont{Nieminen}},
  \bibinfo{journal}{Phys. Rev. B} \textbf{\bibinfo{volume}{51}},
  \bibinfo{pages}{R7341} (\bibinfo{year}{1995}).

\bibitem[{\citenamefont{Barbiellini et~al.}(1996)\citenamefont{Barbiellini,
  Puska, Korhonen, Harju, Torsti, and Nieminen}}]{Barbiellini96}
\bibinfo{author}{\bibfnamefont{B.}~\bibnamefont{Barbiellini}},
  \bibinfo{author}{\bibfnamefont{M.~J.} \bibnamefont{Puska}},
  \bibinfo{author}{\bibfnamefont{T.}~\bibnamefont{Korhonen}},
  \bibinfo{author}{\bibfnamefont{A.}~\bibnamefont{Harju}},
  \bibinfo{author}{\bibfnamefont{T.}~\bibnamefont{Torsti}}, \bibnamefont{and}
  \bibinfo{author}{\bibfnamefont{R.~M.} \bibnamefont{Nieminen}},
  \bibinfo{journal}{Phys Rev. B} \textbf{\bibinfo{volume}{53}},
  \bibinfo{pages}{16201} (\bibinfo{year}{1996}).

\bibitem[{\citenamefont{Arponen and Pajanne}(1979)}]{Arponen79}
\bibinfo{author}{\bibfnamefont{J.}~\bibnamefont{Arponen}} \bibnamefont{and}
  \bibinfo{author}{\bibfnamefont{E.}~\bibnamefont{Pajanne}},
  \bibinfo{journal}{Ann. Phys. (N.Y.)} \textbf{\bibinfo{volume}{121}},
  \bibinfo{pages}{343} (\bibinfo{year}{1979}).

\bibitem[{\citenamefont{Korhonen et~al.}(1996)\citenamefont{Korhonen, Puska,
  and Nieminen}}]{Korhonen96}
\bibinfo{author}{\bibfnamefont{T.}~\bibnamefont{Korhonen}},
  \bibinfo{author}{\bibfnamefont{M.~J.} \bibnamefont{Puska}}, \bibnamefont{and}
  \bibinfo{author}{\bibfnamefont{R.~M.} \bibnamefont{Nieminen}},
  \bibinfo{journal}{Phys.\ Rev.\ B} \textbf{\bibinfo{volume}{54}},
  \bibinfo{pages}{15016} (\bibinfo{year}{1996}).

\bibitem[{\citenamefont{Lantto}(1987)}]{Lantto87}
\bibinfo{author}{\bibfnamefont{L.~J.} \bibnamefont{Lantto}},
  \bibinfo{journal}{Phys. Rev. B} \textbf{\bibinfo{volume}{36}},
  \bibinfo{pages}{5160} (\bibinfo{year}{1987}).

\bibitem[{\citenamefont{Daniuk et~al.}(1987)\citenamefont{Daniuk,
  \surname{Kontrym Sznajd}, Rubaszek, Stachowiak, Mayers, Walters, and
  West}}]{Daniuk87}
\bibinfo{author}{\bibfnamefont{S.}~\bibnamefont{Daniuk}},
  \bibinfo{author}{\bibfnamefont{G.}~\bibnamefont{\surname{Kontrym Sznajd}}},
  \bibinfo{author}{\bibfnamefont{A.}~\bibnamefont{Rubaszek}},
  \bibinfo{author}{\bibfnamefont{H.}~\bibnamefont{Stachowiak}},
  \bibinfo{author}{\bibfnamefont{J.}~\bibnamefont{Mayers}},
  \bibinfo{author}{\bibfnamefont{P.~A.} \bibnamefont{Walters}},
  \bibnamefont{and} \bibinfo{author}{\bibfnamefont{R.~N.} \bibnamefont{West}},
  \bibinfo{journal}{J.\ Phys.\ F} \textbf{\bibinfo{volume}{17}},
  \bibinfo{pages}{1365} (\bibinfo{year}{1987}).

\bibitem[{\citenamefont{Kresse and
  Furthm\"uller}(1996{\natexlab{a}})}]{Kresse96a}
\bibinfo{author}{\bibfnamefont{G.}~\bibnamefont{Kresse}} \bibnamefont{and}
  \bibinfo{author}{\bibfnamefont{J.}~\bibnamefont{Furthm\"uller}},
  \bibinfo{journal}{Comput. Mat. Sci.} \textbf{\bibinfo{volume}{6}},
  \bibinfo{pages}{15} (\bibinfo{year}{1996}{\natexlab{a}}).

\bibitem[{\citenamefont{Kresse and
  Furthm\"uller}(1996{\natexlab{b}})}]{Kresse96b}
\bibinfo{author}{\bibfnamefont{G.}~\bibnamefont{Kresse}} \bibnamefont{and}
  \bibinfo{author}{\bibfnamefont{J.}~\bibnamefont{Furthm\"uller}},
  \bibinfo{journal}{Phys. Rev. B} \textbf{\bibinfo{volume}{54}},
  \bibinfo{pages}{11169} (\bibinfo{year}{1996}{\natexlab{b}}).

\bibitem[{\citenamefont{Kresse and Joubert}(1999)}]{PAWKresse}
\bibinfo{author}{\bibfnamefont{G.}~\bibnamefont{Kresse}} \bibnamefont{and}
  \bibinfo{author}{\bibfnamefont{D.}~\bibnamefont{Joubert}},
  \bibinfo{journal}{Phys. Rev. B} \textbf{\bibinfo{volume}{59}},
  \bibinfo{pages}{1758} (\bibinfo{year}{1999}).

\bibitem[{\citenamefont{Nagai et~al.}(2002)\citenamefont{Nagai, Honma, Tang,
  Hono, and Hasegawa}}]{Nagai02}
\bibinfo{author}{\bibfnamefont{Y.}~\bibnamefont{Nagai}},
  \bibinfo{author}{\bibfnamefont{T.}~\bibnamefont{Honma}},
  \bibinfo{author}{\bibfnamefont{Z.}~\bibnamefont{Tang}},
  \bibinfo{author}{\bibfnamefont{K.}~\bibnamefont{Hono}}, \bibnamefont{and}
  \bibinfo{author}{\bibfnamefont{M.}~\bibnamefont{Hasegawa}},
  \bibinfo{journal}{Phil. Mag. A} \textbf{\bibinfo{volume}{82}},
  \bibinfo{pages}{1559} (\bibinfo{year}{2002}).

\bibitem[{\citenamefont{Nagai et~al.}(2000)\citenamefont{Nagai, Tang, and
  Hasegawa}}]{Nagai00}
\bibinfo{author}{\bibfnamefont{Y.}~\bibnamefont{Nagai}},
  \bibinfo{author}{\bibfnamefont{Z.}~\bibnamefont{Tang}}, \bibnamefont{and}
  \bibinfo{author}{\bibfnamefont{M.}~\bibnamefont{Hasegawa}},
  \bibinfo{journal}{Rad.\ Phys. Chem.} \textbf{\bibinfo{volume}{58}},
  \bibinfo{pages}{737} (\bibinfo{year}{2000}).

\bibitem[{\citenamefont{Monkhorst and Pack}(1976)}]{Monkhorst76}
\bibinfo{author}{\bibfnamefont{H.~J.} \bibnamefont{Monkhorst}}
  \bibnamefont{and} \bibinfo{author}{\bibfnamefont{J.~D.} \bibnamefont{Pack}},
  \bibinfo{journal}{Phys. Rev. B} \textbf{\bibinfo{volume}{13}},
  \bibinfo{pages}{5188} (\bibinfo{year}{1976}).

\bibitem[{\citenamefont{Schaefer et~al.}(1984)\citenamefont{Schaefer,
  Gugelmeier, Schmolz, and Seeger}}]{Schaefer84}
\bibinfo{author}{\bibfnamefont{H.~E.} \bibnamefont{Schaefer}},
  \bibinfo{author}{\bibfnamefont{R.}~\bibnamefont{Gugelmeier}},
  \bibinfo{author}{\bibfnamefont{M.}~\bibnamefont{Schmolz}}, \bibnamefont{and}
  \bibinfo{author}{\bibfnamefont{A.}~\bibnamefont{Seeger}}, in
  \emph{\bibinfo{booktitle}{Proceedings of the Vth Ris\o\ International
  Symposium of Metallurgy and Materials Science}}, edited by
  \bibinfo{editor}{\bibfnamefont{N.~H.} \bibnamefont{Andersen}},
  \bibinfo{editor}{\bibfnamefont{M.}~\bibnamefont{Eldrup}},
  \bibinfo{editor}{\bibfnamefont{N.}~\bibnamefont{Hansen}},
  \bibinfo{editor}{\bibfnamefont{D.~J.} \bibnamefont{Jensen}},
  \bibinfo{editor}{\bibfnamefont{T.}~\bibnamefont{Leffers}},
  \bibinfo{editor}{\bibfnamefont{H.}~\bibnamefont{Lillholt}},
  \bibinfo{editor}{\bibfnamefont{O.~B.} \bibnamefont{Pedersen}},
  \bibnamefont{and} \bibinfo{editor}{\bibfnamefont{B.~N.} \bibnamefont{Singh}}
  (\bibinfo{publisher}{Ris\o\ National Laboratory}, \bibinfo{address}{Ris\o},
  \bibinfo{year}{1984}).

\bibitem[{\citenamefont{Schaefer et~al.}(1986)\citenamefont{Schaefer, Stuck,
  Banhart, and Bauer}}]{Schaefer86}
\bibinfo{author}{\bibfnamefont{H.~E.} \bibnamefont{Schaefer}},
  \bibinfo{author}{\bibfnamefont{W.}~\bibnamefont{Stuck}},
  \bibinfo{author}{\bibfnamefont{F.}~\bibnamefont{Banhart}}, \bibnamefont{and}
  \bibinfo{author}{\bibfnamefont{W.}~\bibnamefont{Bauer}}, in
  \emph{\bibinfo{booktitle}{Proceedings of the 8th International Conference on
  Vacancies and Interstitials in Metals and Alloys}}, edited by
  \bibinfo{editor}{\bibfnamefont{C.}~\bibnamefont{Ambromeit}} \bibnamefont{and}
  \bibinfo{editor}{\bibfnamefont{H.}~\bibnamefont{Wollenberger}}
  (\bibinfo{publisher}{Trans Tech}, \bibinfo{address}{Aedermannsdorf},
  \bibinfo{year}{1986}).

\bibitem[{\citenamefont{Laine et~al.}(1996)\citenamefont{Laine, Saarinen,
  M\"akinen, Hautoj\"arvi, Corbel, Pfeiffer, and Citrin}}]{Laine96}
\bibinfo{author}{\bibfnamefont{T.}~\bibnamefont{Laine}},
  \bibinfo{author}{\bibfnamefont{K.}~\bibnamefont{Saarinen}},
  \bibinfo{author}{\bibfnamefont{J.}~\bibnamefont{M\"akinen}},
  \bibinfo{author}{\bibfnamefont{P.}~\bibnamefont{Hautoj\"arvi}},
  \bibinfo{author}{\bibfnamefont{C.}~\bibnamefont{Corbel}},
  \bibinfo{author}{\bibfnamefont{L.~N.} \bibnamefont{Pfeiffer}},
  \bibnamefont{and} \bibinfo{author}{\bibfnamefont{P.~H.}
  \bibnamefont{Citrin}}, \bibinfo{journal}{Phys.\ Rev. B}
  \textbf{\bibinfo{volume}{54}}, \bibinfo{pages}{R11050}
  (\bibinfo{year}{1996}).

\bibitem[{\citenamefont{Corbel et~al.}(1992)\citenamefont{Corbel, Pierre,
  Saarinen, Hautoj\"arvi, and Moser}}]{Corbel92}
\bibinfo{author}{\bibfnamefont{C.}~\bibnamefont{Corbel}},
  \bibinfo{author}{\bibfnamefont{F.}~\bibnamefont{Pierre}},
  \bibinfo{author}{\bibfnamefont{K.}~\bibnamefont{Saarinen}},
  \bibinfo{author}{\bibfnamefont{P.}~\bibnamefont{Hautoj\"arvi}},
  \bibnamefont{and} \bibinfo{author}{\bibfnamefont{P.}~\bibnamefont{Moser}},
  \bibinfo{journal}{Phys.\ Rev.\ B} \textbf{\bibinfo{volume}{45}},
  \bibinfo{pages}{3386} (\bibinfo{year}{1992}).

\bibitem[{\citenamefont{Puska et~al.}(1998)\citenamefont{Puska, P\"oykk\"o,
  Pesola, and Nieminen}}]{Puska98}
\bibinfo{author}{\bibfnamefont{M.~J.} \bibnamefont{Puska}},
  \bibinfo{author}{\bibfnamefont{S.}~\bibnamefont{P\"oykk\"o}},
  \bibinfo{author}{\bibfnamefont{M.}~\bibnamefont{Pesola}}, \bibnamefont{and}
  \bibinfo{author}{\bibfnamefont{R.~M.} \bibnamefont{Nieminen}},
  \bibinfo{journal}{Phys.\ Rev.\ B} \textbf{\bibinfo{volume}{58}},
  \bibinfo{pages}{1318} (\bibinfo{year}{1998}).

\bibitem[{\citenamefont{Probert and Payne}(2003)}]{Probert03}
\bibinfo{author}{\bibfnamefont{M.~I.~J.} \bibnamefont{Probert}}
  \bibnamefont{and} \bibinfo{author}{\bibfnamefont{M.~C.} \bibnamefont{Payne}},
  \bibinfo{journal}{Phys.\ Rev.\ B} \textbf{\bibinfo{volume}{67}},
  \bibinfo{pages}{075204} (\bibinfo{year}{2003}).

\bibitem[{\citenamefont{Latham et~al.}(2005)\citenamefont{Latham, Ganchenkova,
  Nieminen, Nicolaysen, Alatalo, \"Oberg, and Briddon}}]{Latham05}
\bibinfo{author}{\bibfnamefont{C.~D.} \bibnamefont{Latham}},
  \bibinfo{author}{\bibfnamefont{M.}~\bibnamefont{Ganchenkova}},
  \bibinfo{author}{\bibfnamefont{R.~M.} \bibnamefont{Nieminen}},
  \bibinfo{author}{\bibfnamefont{S.}~\bibnamefont{Nicolaysen}},
  \bibinfo{author}{\bibfnamefont{M.}~\bibnamefont{Alatalo}},
  \bibinfo{author}{\bibfnamefont{S.}~\bibnamefont{\"Oberg}}, \bibnamefont{and}
  \bibinfo{author}{\bibfnamefont{P.~R.} \bibnamefont{Briddon}},
  \bibinfo{journal}{Physica Scripta, submitted}  (\bibinfo{year}{2005}).

\bibitem[{\citenamefont{Makhov and Lewis}(2005)}]{Makhov05}
\bibinfo{author}{\bibfnamefont{D.~V.} \bibnamefont{Makhov}} \bibnamefont{and}
  \bibinfo{author}{\bibfnamefont{L.~J.} \bibnamefont{Lewis}},
  \bibinfo{journal}{Phys.\ Rev.\ B} \textbf{\bibinfo{volume}{71}},
  \bibinfo{pages}{205215} (\bibinfo{year}{2005}).

\bibitem[{\citenamefont{M\"akinen et~al.}(1989)\citenamefont{M\"akinen, Corbel,
  Hautoj\"arvi, Moser, and Pierre}}]{Makinen89}
\bibinfo{author}{\bibfnamefont{J.}~\bibnamefont{M\"akinen}},
  \bibinfo{author}{\bibfnamefont{C.}~\bibnamefont{Corbel}},
  \bibinfo{author}{\bibfnamefont{P.}~\bibnamefont{Hautoj\"arvi}},
  \bibinfo{author}{\bibfnamefont{P.}~\bibnamefont{Moser}}, \bibnamefont{and}
  \bibinfo{author}{\bibfnamefont{F.}~\bibnamefont{Pierre}},
  \bibinfo{journal}{Phys. Rev. B} \textbf{\bibinfo{volume}{39}},
  \bibinfo{pages}{10162} (\bibinfo{year}{1989}).

\bibitem[{\citenamefont{Polity et~al.}(1998)\citenamefont{Polity, B\"orner,
  Huth, Eichler, and Krause-Rehberg}}]{Polity98}
\bibinfo{author}{\bibfnamefont{A.}~\bibnamefont{Polity}},
  \bibinfo{author}{\bibfnamefont{F.}~\bibnamefont{B\"orner}},
  \bibinfo{author}{\bibfnamefont{S.}~\bibnamefont{Huth}},
  \bibinfo{author}{\bibfnamefont{S.}~\bibnamefont{Eichler}}, \bibnamefont{and}
  \bibinfo{author}{\bibfnamefont{R.}~\bibnamefont{Krause-Rehberg}},
  \bibinfo{journal}{Phys.\ Rev.\ B} \textbf{\bibinfo{volume}{58}},
  \bibinfo{pages}{10363} (\bibinfo{year}{1998}).

\bibitem[{\citenamefont{Saito and Oshiyama}(1996)}]{Saito96}
\bibinfo{author}{\bibfnamefont{M.}~\bibnamefont{Saito}} \bibnamefont{and}
  \bibinfo{author}{\bibfnamefont{A.}~\bibnamefont{Oshiyama}},
  \bibinfo{journal}{Phys.\ Rev.\ B} \textbf{\bibinfo{volume}{53}},
  \bibinfo{pages}{7810} (\bibinfo{year}{1996}).

\bibitem[{\citenamefont{El-Mellouhi and Mousseau}(2005)}]{El-Mellouhi05}
\bibinfo{author}{\bibfnamefont{F.}~\bibnamefont{El-Mellouhi}} \bibnamefont{and}
  \bibinfo{author}{\bibfnamefont{N.}~\bibnamefont{Mousseau}},
  \bibinfo{journal}{Phys.\ Rev.\ B} \textbf{\bibinfo{volume}{71}},
  \bibinfo{pages}{125207} (\bibinfo{year}{2005}).

\end{thebibliography}

\end{document}